\documentclass[runningheads]{comsis2}

\def\journalissue{Computer Science and Information Systems 00(0):0000--0000}
\def\paperidnum{https://doi.org/10.2298/CSIS123456789X}
\usepackage{dutchcal}
\usepackage[pagewise]{lineno}
\usepackage{algorithm}
\usepackage[noend]{algpseudocode}
\usepackage[T1]{fontenc}
\usepackage{amsmath}
\usepackage{hyperref,subcaption}
\usepackage{verbatim}
\usepackage{graphicx}
\usepackage{booktabs,nicematrix,pict2e,braket,xfrac}
\usepackage{glossaries}
\usepackage{enumitem}
\DeclareMathOperator{\dom}{dom}
\newlist{inline}{enumerate*}{1}
\setlist[inline]{label=\textit{(\roman*)}}
\makeglossaries
\newacronym{bpm}{BPM}{Business Process Management}
\newacronym{dt}{DT}{Data Trends}
\newacronym{ts}{TS}{Time Series}
\newacronym{mts}{MTS}{Multivariate Time Series}
\newacronym{mtsc}{MTSC}{Multivariate Time Series Classification}
\newacronym{dag}{DAG}{Direct Acyclic Graph}
\newacronym{llm}{LLM}{Large Language Model}
\newacronym{emeritate}{EMeriTAte}{Explainable MultivariatE coRrelatIonal Temporal Artificial inTElligence}
\newacronym{emeritatedf}{EMeriTAte+DF}{DataFul Explainable MultivariatE coRrelatIonal Temporal Artificial inTElligence}

\usepackage{bm,amsfonts,amsbsy,amsmath,stmaryrd}
\title{Towards Explainable Sequential Learning%
\footnote{This journal paper constitutes a major extension of our previous work \cite{ideas2024a}.}}

\titlerunning{Towards Explainable Sequential Learning}
\author{Giacomo Bergami\inst{1} \and Emma Packer \inst{2} \and Kirsty Scott \inst{2} \and Silvia Del Din \inst{2,3}} 
\authorrunning{Bergami et al.}
\institute{School of Computing, Newcastle University\\
  \email{Giacomo.Bergami@newcastle.ac.uk}
  \and
Translational and Clinical Research Institute, Faculty of Medical Sciences,
Newcastle University\\
  \email{\{e.packer,kirsty.scott-singer\}@newcastle.ac.uk}
  \and
National Institute for Health and Care Research (NIHR), Newcastle Biomedical
Research Centre (BRC), Newcastle University and The Newcastle upon Tyne
Hospitals NHS Foundation Trust \\
  \email{silvia.del-din@newcastle.ac.uk}
}

\usepackage{arydshln}
\usepackage{inconsolata}

\usepackage{setspace}
\newcommand{\yieldconst}[6]{\textbf{yield} $\varsigma^\mathfrak{E}_{#1,\texttt{fresh}(#1)}:=\braket{\textsc{DL}(#6,#4;x,i), \textsc{C22}(#5[#1,#2])\circ \textsc{C22}([#1,#2]),#3}$}
\newcommand{\DeclareClauseNoData}[3]{\textsf{#1}(\texttt{#2},\texttt{#3})}

\begin{document}

\maketitle

\begin{abstract}

This paper offers a hybrid explainable temporal data processing pipeline, \gls{emeritatedf}, bridging numerical-driven temporal data classification with an event-based one through verified artificial intelligence principles, enabling human-explainable results. This was possible through a preliminary \textit{a posteriori} explainable phase describing the numerical input data in terms of concurrent constituents with numerical payloads. This further required extending the event-based literature to design specification mining algorithms supporting concurrent constituents. Our previous and current solutions outperform state-of-the-art solutions for multivariate time series classifications, thus showcasing the effectiveness of the proposed methodology.

\vspace{6pt}\textbf{Keywords:} Verified AI; eXplainable AI (XAI); polyadic logs; Data Trends; Poly-DECLARE; Multivariate Time Series Classification
\end{abstract}

\newcommand{\fromhere}{\color{red}\P}

\section{Introduction}

The current literature on temporal data analysis is divided into two main branches: one, mainly considering numerical fluctuations in different dimensions as \gls{mts} \cite{DempsterPW20,9378424,Hills2014,ZhangG0L20}, and the other focusing on a characterization as a linear sequence of discrete events, be they durative\cite{Allen,RostGTFSCAJR22,5963680} or pointwise \cite{XuLZ17a,bolt2}. The first class denotes the classification task as \gls{mtsc}, which requires capturing behavioural correlations between different dimensions \cite{Ruiz2021}. Cross-dimension correlations can be represented via convolutions \cite{ZhangG0L20} or kernels that potentially squash such correlations to a single numerical value \cite{DempsterPW20}. Despite the possibility of doing so, the explanations thus obtained are not easily explainable to humans. Furthermore, recent results have shown that these solutions cannot capture such correlations when considering complex and multifactorial problems, such as clinical ones, and give results with low accuracy \cite{ideas2024a}. These considerations require us to solve the problem from a different angle.

Recent approaches are starting to bridge the numerical with the event-driven characterization of temporal data by clearly showing the possibility of discretizing \gls{dt} occurring into durative constituents \cite{HUO2022117176}. Notwithstanding this, these events cannot summarize numerical information precisely characterizing the trend of choice, thus only characterizing growth, variability, or decrease patterns without providing any further numerical references to better describe the event of choice. This demands a more general representation of DT patterns, which should also encompass numerical features, albeit summarized (e.g. Catch22 \cite{catch24}), to characterize the underlying data from a more data-driven perspective.

Within the second branch, Business Process Mining \cite{vanderAalst2022,XuLZ17a} enables, through declarative process mining, to temporally correlate single pointwise events through a human-readable data description while bridging across different \textit{activity labels} (distinguishing different action/event types) occurring. This literature premiered the extraction of declarative specifications from event-based data while enriching information concerning temporal correlations with additional numerical \cite{LENO2020101482} and categorical \cite{Undone} data predicates. Despite the inherent ability of such techniques to provide a human-explainable characterization of the data, the former techniques cannot be straightforwardly applied to solve the \gls{mtsc} problem, even after \gls{dt}-discretization. \textit{First}, despite the possibility of representing temporal specifications as concurrent ones \cite{vanderAalst2022}, these are always interpreted as a temporally-ordered succession of non-concurrent and non-overlapping events \cite{9576856}, thus invalidating the possibility of directly exploiting the same algorithms for capturing multiple events co-occurring at the same time across dimensions. In fact, despite the possibility for the current XES \cite{XES} standard to express durative events, this cannot clearly represent and group together co-occurring events within the same trace across dimensions. \textit{Second}, as such techniques often involve small and curated datasets, they have been proven several times as not scalable over real-world ones \cite{bolt2}. At the time of the writing, the most efficient temporal specification miner, Bolt2 \cite{bolt2}, extracts data-less temporal clauses and, therefore, does not consider data payload information associated with each activity label, as required after extending \gls{dt} with data payloads. As this was the one used in our previous work to characterize correlations across temporal dimensions after a minor extension for supporting durative and co-occurring events \cite{ideas2024a}, the extraction of a declarative dataful specification requires to expand the latter algorithm further also to enable the extraction of dataful features from the declarative clauses. 

This paper aims to bridge all approaches above to achieve a generally explainable sequential learning explanation through \gls{emeritatedf}\footnote{\url{https://github.com/datagram-db/knobab/releases/tag/v3.1}}. The proposed technique works by first discretizing \gls{mts} into machine-readable polyadic logs through dataful \gls{dt} mining (Sect. \ref{dtmining}), for then extracting declarative descriptions of the previously-mined representation while pertaining dataful features (Sect. \ref{sec:polymine}). We then use such clauses to derive the features satisfied, violated, or vacuously satisfied by the traces, for then deriving a propositional representation of the classes associated with arbitrary \gls{mts} segments through white-box classifier explanation extraction (also Sect. \ref{sec:polymine}). By doing so, each temporal class is then defined through the satisfaction of a proportional formula, where predicates state temporal properties of the data. This approach further generalises pre-existing deviance mining solutions, which only encompass the conjunctive-based characterization of such temporal classes \cite{Undone}.

This constitutes a major improvement over our previous solution \gls{emeritate} \cite{ideas2024a}, where \begin{inline}
\item the original implementation of DT mining was not extracting a numerical summarised description of the data and \item was using a straightforward brute-force algorithm for considering all the possible sliding windows of all possible sizes; \item the extraction of the declarative clauses for each polyadic log (Sect. \ref{sec:pollog}) did not consider dataful clauses, and \item the generation of the embedding from the mined clauses was regarded as a distinct phase from the one where the trace embeddings were generated as in previous literature \cite{Undone}, thus adding significant computational overhead. 
\end{inline} The application of the aforementioned changes to our previous solution enables us to design its dataful variant (\gls{emeritatedf}) outperforming \gls{emeritate} in terms of efficiency, accuracy, precision, and recall. Last, we extend our experimental set-up to consider other \gls{mts} datasets, thus including univariate ones \cite{italypowerdemand,osuleaf} and others related to general human mobility, thus remarking the inherent difficulty of characterizing dyskinetic/off events in Parkinson's Disease patients.

\section{Related Works}


\subsection{Hybridly Explainable Artificial Intelligence}
The observations from which the utility of defining a hybrid explainable AI is deduced have their origins in the observations deduced from the formalization of verifiable AI systems. At the basis of this framework \cite{10.1145/3503914}, we always consider a model of the system to be verified ($\mathfrak{S}$), a model of the environment ($\mathfrak{E}$), and the formal property to be verified ($\varphi$), where all of these components might be expressed in logical form, so to provide correctness guarantees. Specification mining algorithms \cite{bolt2} extract $\varphi$ from given system or environment of interest $\mathfrak{S}\vee\mathfrak{E}$, where deviance mining algorithms \cite{Undone} are a specific case of the former, assuming  \begin{inline}
\item that environment states can be labelled and
\item that we can extract one single specification per environment label, which provides the characterizing behaviour distinguishing a class from the others.
\end{inline}. A later survey \cite{chapter} generalised the former into \textit{hybridly explainable artificial intelligence} after observing that we can always transform unstructured data representations into more structured and logical-driven by pre-processing the data into an \textit{a priori} phase, which can include both data cleaning capabilities \cite{wrembel} and specification mining ones by enriching structured data representation with contextual interpretations of the raw data. After this, we can carry out the actual learning phase, in which \textit{ad hoc} explainability will heavily depend on the methodology and the data-representation format of choice. Last, we can consider precision/recall metrics as well as model extractions from both white-box and black-box classifiers as \textit{ex post} explanations of the learning phase. This paper will focus on providing algorithmic solutions for the first two phases, as narrated in the text.

To better explain the former framework, we will categorize the forthcoming sections by pigeonholing curring literature over the aforementioned framework.

\subsection{A Priori Explanability}

\begin{figure}[!p]

\begin{subfigure}{\textwidth}
  \centering
\resizebox{.7\linewidth}{!}{\begin{tabular}{l|cl}
			\toprule
			  Activity Label  &  \gls{ts} & Trace ($A^\texttt{i},A^{\neg\texttt{i}}$)\\
			\midrule
	  \textsf{IncreaseRapidly(A\textsuperscript{\texttt{i}})}  & \includegraphics[height=.5cm]{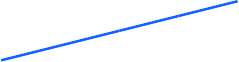} & $A^\texttt{i},\dots,A^\texttt{i}$\\
	  \textsf{IncreaseSlowlyI(A\textsuperscript{\texttt{i}})}  & \includegraphics[height=.5cm]{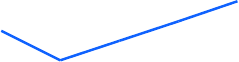} & $A^{\neg\texttt{i}},A^\texttt{i},\dots,A^\texttt{i}$\\
	  \textsf{IncreaseSlowlyII(A\textsuperscript{\texttt{i}})}  & \includegraphics[height=.5cm]{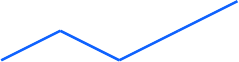} & $A^\texttt{i},A^{\neg\texttt{i}},A^\texttt{i},\dots,A^\texttt{i}$\\
	  \textsf{IncreaseSlowlyIII(A\textsuperscript{\texttt{i}})}  & \includegraphics[height=.5cm]{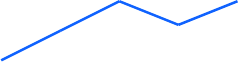} & $A^\texttt{i},\dots,A^\texttt{i},A^{\neg\texttt{i}},A^\texttt{i}$\\
	  \textsf{IncreaseSlowlyIV(A\textsuperscript{\texttt{i}})}  & \includegraphics[height=.5cm]{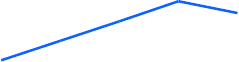} & $A^\texttt{i},\dots,A^\texttt{i},A^{\neg\texttt{i}}$\\
	  \textsf{HighVolatilityI(A\textsuperscript{\texttt{i}})}  & \includegraphics[height=.5cm]{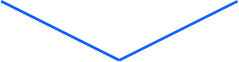} & $A^{\neg\texttt{i}},A^\texttt{i}$\\
	  \textsf{HighVolatilityII(A\textsuperscript{\texttt{i}})}  & \includegraphics[height=.5cm]{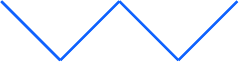} & $A^{\neg\texttt{i}},A^{\texttt{i}},A^{\neg\texttt{i}},A^{\texttt{i}}$\\
	  \textsf{HighVolatilityIII(A\textsuperscript{\texttt{i}})}  & \includegraphics[height=.5cm]{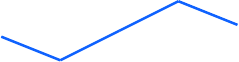}& $A^{\neg\texttt{i}},A^\texttt{i},\dots,A^\texttt{i},A^{\neg\texttt{i}}$\\
	  \textsf{HighVolatilityIV(A\textsuperscript{\texttt{i}})}  & \includegraphics[height=.5cm]{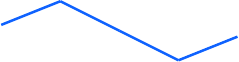}& $A^{\texttt{i}},A^{\neg\texttt{i}},\dots,A^{\neg\texttt{i}},A^{\texttt{i}}$\\
	  \textsf{HighVolatilityV(A\textsuperscript{\texttt{i}})}  & \includegraphics[height=.5cm]{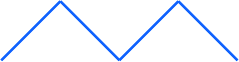}& $A^{\texttt{i}},A^{\neg\texttt{i}},A^{\texttt{i}},A^{\neg\texttt{i}}$\\
	  \textsf{HighVolatilityVI(A\textsuperscript{\texttt{i}})}  & \includegraphics[height=.5cm]{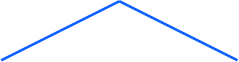} & $A^{\texttt{i}},A^{\neg\texttt{i}}$\\
\textsf{DecreaseSlowlyI(A\textsuperscript{\texttt{i}})}  & \includegraphics[height=.5cm]{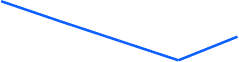}& $A^{\neg\texttt{i}},\dots,A^{\neg\texttt{i}},A^{\texttt{i}}$\\
	  \textsf{DecreaseSlowlyII(A\textsuperscript{\texttt{i}})}  & \includegraphics[height=.5cm]{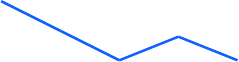}& $A^{\neg\texttt{i}},\dots,A^{\neg\texttt{i}},A^{\texttt{i}},A^{\neg\texttt{i}}$\\
	  \textsf{DecreaseSlowlyIII(A\textsuperscript{\texttt{i}})}  & \includegraphics[height=.5cm]{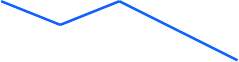}& $A^{\neg\texttt{i}},A^{\texttt{i}},A^{\neg\texttt{i}},\dots,A^{\neg\texttt{i}}$\\
	  \textsf{DecreaseSlowlyIV(A\textsuperscript{\texttt{i}})}  & \includegraphics[height=.5cm]{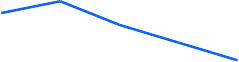}& $A^{\texttt{i}},A^{\neg\texttt{i}},\dots,A^{\neg\texttt{i}}$\\
	  \textsf{DecreaseRapidly(A\textsuperscript{\texttt{i}})}  & \includegraphics[height=.5cm]{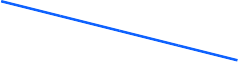}& $A^{\neg\texttt{i}},\dots,A^{\neg\texttt{i}}$\\
			\bottomrule
	\end{tabular}\qquad}
  \caption{Expressing \gls{dt} activity labels as \gls{ts} silhouettes and pointwise constituents \cite{HUO2022117176}.}
  \label{fig:freighttrain}
\end{subfigure}
\begin{subfigure}{.7\textwidth}
\resizebox{1.3\linewidth}{!}{\begin{tabular}{lp{8.5cm}}
			\toprule
			  Template ($c_l$)  & Non-polyadic logs interpretation $\llbracket c_l\rrbracket$\\
			\midrule
            \textdagger\textsf{Init($A,p$)} & The trace should start with an activation\\
                        \textdagger\textsf{End($A,p$)} & The trace should end with an activation\\
	  \textdagger\textsf{Exists($A,p$)}  & Activations should occur at least once \\
	  \textsf{Absence($A,p$)}  & Activations should never occur\\
	 \textsf{Choice($A,p,A',p'$) }  & One of the two activation  conditions must appear. \\
	  \textsf{ExclChoice($A,p,A',p'$) }   & Only one type of activation is admitted per trace.\\ 
	 \textsf{RespExistence($A,p,B,q$) }   & An activation requires at least one target.  \\
	  \textsf{CoExistence($A,p,B,q$) }    & $ \llbracket\DeclareClauseNoData{RespExistence}{A}{B}\rrbracket\wedge \llbracket\DeclareClauseNoData{RespExistence}{B}{A}\rrbracket$\\
	  \textdagger\textsf{Precedence($A,p,B,q$)} & Events preceding the activations should not satisfy the target.\\
	  \textdagger\textsf{Response($A,p,B,q$) }  & An activation leads to a future target event. \\
	  \textsf{Succession($A,B$) }  & 
 $\llbracket\DeclareClauseNoData{Precedence}{A}{B}\rrbracket\wedge \llbracket\DeclareClauseNoData{Response}{A}{B}\rrbracket$\\
				 \textdagger\textsf{ChainPrecedence($A,p,B,q$) }   & An activation leads to an \textit{immediately preceding} target. \\ 
		 \textdagger\textsf{ChainResponse($A,p,B,q$) }   & An activation leads to an \textit{immediately following} target.\\
	  \textsf{ChainSuccession($A,p,B,q$) }  &   $\llbracket\DeclareClauseNoData{ChainPreced.}{B}{A}\rrbracket\wedge \llbracket\DeclareClauseNoData{ChainResponse}{A}{B}\rrbracket$\\
			\bottomrule
	\end{tabular}}
  \caption{Our dataful DECLARE \cite{PesicSA07} subset of interest, where $A$ {(respectively, $B$) denote activation (resp., target), and $p$ (resp., $q$) denote the associated dataful payload condition.} conditions. Dataless variants can be expressed with $p$ ($p'$ and $q$) as \textbf{true}. \textdagger remarks clauses subject to dataful refinement in the present paper.}
  \label{tab:dt}
\end{subfigure}
\caption{Declarative Languages for \textit{a priori} explanability over \textit{(a)} \gls{ts} or \textit{(b)} non-polyadic logs.}
\label{fig:runsofast}
\end{figure}
\subsubsection{Numerical Data.} \textsc{\gls{dt}} specifications (\figurename~\ref{fig:freighttrain}) provide a LTL\textsubscript{f} characterization for
 numerical trends in (univariate) \gls{ts} $X$ after propositional discretization within a mining phase \cite{HUO2022117176}: this requires a preliminary pre-processing step for which $X$ is associated with durative constituents $\varsigma^X_{j+1}$ with activity labels $X^\texttt{c}$ or $X^{\neg\texttt{c}}$: $\varsigma^X_{j+1}$ has activity label $X^\texttt{c}$ ($\lambda(\varsigma^x_{j+1})=X^\texttt{c}$) if $X(j+1)>X(t)$ and $X^{\neg\texttt{c}}$ otherwise. We can then determine the occurrence of the pattern by simply joining constituents $X^\texttt{c}$ and $X^{\neg\texttt{c}}$ into durative ones referring to the same temporal span, which can then be matched and rewritten into a durative event having the activity label described in the first column of \figurename~\ref{fig:freighttrain}. Authors mainly use this declarative representation for \gls{ts} forecasting purposes, but give no evidence for exploiting this in the context of \gls{mts} while correlating disparate \gls{dt} across \gls{mts} variables, which might be expressed through DECLARE$d$. To achieve this, our previous work premiered their combination. Although \gls{dt} discards the numerical information associated with such trends, it generalises the common shapelet approach \cite{Hills2014} to arbitrary growth and decrease patterns expressed declaratively with human-understandable  patterns.

\subsubsection{Event Data.} \textsc{DECLARE} \cite{PesicSA07}  (\figurename~\ref{tab:dt}) is a \textit{log} temporal declarative language considering possible temporal behaviour in terms of correlations between activated events and targeted conditions, where the former refers to the necessary but not sufficient condition for satisfying a clause. Both activation and target conditions are defined through \textit{activity labels} referring to a specific action of interest; these can be enriched with dataful propositional formul\ae, predicating payload conditions over the activated constituents. Unless stated otherwise, if the clause has no activation, it is trivially satisfied (\textit{vacuously}). E.g., \textsf{Precedence} might also be vacuously violated due to the presence of just target conditions \cite{bolt2}. 
DECLARE clauses can be instantiated over a specific alphabet $\Sigma$ referring to activity labels associated with constituents and they can be composed to generate \textit{conjunctive specifications} $\Phi=\{c_1,\dots,c_l\}$; we say that a trace satisfies a conjunctive specification if it satisfies all its clauses. These conjunctive specifications can be extracted from temporal data represented as a (non-polyadic) log by 
efficiently exploiting a specific-to-general lattice search from which we can preventively prune the generation of specific clauses \cite{bolt2}. 

 Bolt2 \cite{bolt2} mines non-polyadic logs to extract conjunctive DECLARE specifications, where events are non-durative and where every single event contains one and only one constituent. This algorithm achieves efficient specification mining by testing DECLARE dataless clauses where the most general clause is tested first, followed by the ones enabling the other. This generates a lattice of DECLARE clauses: the clause search stops by adopting a quality metric, thus ensuring the capture of the most specific behaviour of the single log without returning underrepresented ones as not generalizing over the data. This solution does not contemplate any further refinement of the clauses into dataful ones so to associate non-trivially true data predicates to either the activation, target, or correlation conditions. Other specification mining algorithms overcome this limitation \cite{LENO2020101482} and, despite not using an efficient general-to-specific visit as the former, consider clustering over the activation conditions for deriving the data's propositional representations. This consideration was blindly adopted in DML \cite{Undone} to refine each mined declare clause to generate dataful clauses differentiating traces belonging to different classes. This poses a significant limitation in this scenario, as the mining approach was not initially designed for this purpose, thus not necessarily helping to refine the activation conditions according to the class associated with each single activation or target condition. This paper proposes a completely different approach, where white-box classifiers are used to characterize the activation (and target) predicates with the clear intent to capture distinct class behaviours and not merely aggregate data by shared data features, which might not be necessarily related to a class of interest.

\subsection{Ad Hoc Explanability}\label{competingApproaches}

\subsubsection{Numerical Data.} 
\textit{This subsection describes the main competitors considered from this paper for \gls{emeritatedf}, as they constitute \gls{mts} classifiers.}

{(KNN-based) Clustering for time series \cite{Inoue2024} works by first identifying time series clusters depending on a distance function of choice and then associating to each cluster the majority class; this induces the possibility of boiling down a clustering problem to a classification one by associating each numerical-based representation to the element being the most similar to the majority class similarly to \cite{AvolioFVZ23}. A straightforward Euclidean metric can determine the pointwise distance of the \gls{mts} across dimensions (E-KNN). The main drawback of this is that it neither considers the evolution of \glspl{dt} within the \gls{mts} nor allows for aligning similar trends being displaced in time, similar to dynamic time warping. Furthermore, no efficient mechanisms are designed to efficiently test different patterns at stake by considering a hierarchy of temporal patterns, one being the generalization of the other. Doing so can greatly boost the mining task while increasing the overall number of temporal patterns of interest.

Rocket \cite{DempsterPW20} exploits randomized convolutional kernels to extract relevant features from the \gls{mts}, which are then fed to a linear classifier by associating such features to a class of interest. While this feature extraction approach cannot adequately capture \glspl{dt} and variations across dimensions, it also guarantees scarce explainability as information is summarised into kernel values. TapNet \cite{ZhangG0L20} improves the latter by exploiting attention networks for selecting the best classification features while still heavily relying on up-front feature selection mechanisms such as dimensionality reduction, which might lose relevant information to the detriment of the classification precision and accuracy.
\textsc{Canonical Interval Forest} ({CIF}) \cite{9378424} achieves explainability by exploiting white-box classifiers such as decision trees over \textit{catch22} features \cite{catch24} describing a selection of time intervals describing different types of numerical variations rather than changes in trends and data patterns and their temporal correlations.
\textsc{Shape\-let Transform Classifier} ({STC}) \cite{Hills2014} characterize \gls{mts} in terms of distinctive temporal features, the \textit{shapelets}, being frequently occurring trends occurring across all \gls{ts}, for then describing each \gls{mts} in terms of their distance with each selected shapelet. Notwithstanding this approach considers numerical features that are completely discarded through \gls{dt}, it fails to establish correlations across trends occurring in different dimensions, thus not establishing correlations across differently observed behaviours.

\subsubsection{Event Data.} \textsc{Deviant Model Learning (DML)} \cite{Undone} extracts specifications for trace classes $y\in\mathcal
{Y}$  through declarative (e.g., DECLARE clauses) and procedural (i.e., association rules) features previously mined from the data. Each trace can be represented as an $n$-dimensional embedding $\vec{x}_\sigma\in\mathbb{R}^n$, where $n$ is the overall number of features. E.g., when a Declare clause $c_\ell$ is chosen as the $\ell$-th feature, a value of $\vec{x}_\sigma[\ell]=-1/0/n $   denotes that the trace $\sigma$ does not satisfy, or vacuously satisfies, or satisfactorily activates $c_\ell$ $n$ times. For the features providing declarative characterizations, DML extracts those via the ADMS specification mining algorithm \cite{bolt2}. 
Our previous approach, \gls{emeritate} \cite{ideas2024a}, generalised over the former algorithm in its \textit{ad hoc} phase by considering a different log model, where each event (and not just a single trace) is associated with a class and where each event is composed by different \textit{constituents} representing concurrent activities starting at the same time and with different durations. These characteristics also enable the Event Data algorithms to consider numerical-driven problems with different dimensions. At the same time, previous temporal declarative-driven approaches were limited to only considering univariate \gls{ts} \cite{donze13}, and were only used for verification tasks, but not for specification mining ones. We further expand on this data model to consider multiple taxonomies of activity labels for considering which events might be regarded as sub-events of one more general type (Sect. \ref{sec:pollog}), so as to reduce better the amount of relevant correlations (Sect. \ref{sec:polymine}).

\section{Problem Statement}
We denote $\mathbb{B}(b)$ as the function mapping true values for $b$ to $1$ and returning $-1$ otherwise.

Given a function $F$ having as a domain a closed interval in $\mathbb{N}$, the set $\mathbb{I}_F([B,E])$ of the maximal intervals in $[B,E]\subseteq\mathbb{N}$ is a set of largest non-overlapping intervals in $[B,E]$ sharing the same value for $F$ where all the remaining intervals will have different values in $F$:

\[\begin{split}
\mathbb{I}_F([B,E])=\{[b,e]\,|\,&B\leq b\leq e\leq E, (b>B\Rightarrow F(b)\neq F(B)),\\
                                &(e<E\Rightarrow F(e)\neq F(E)), \forall b\leq \tau\leq e. F(\tau)=F(b)=F(e)\}
\end{split}\]

Given a \gls{mts} $T$, we refer to its \textit{size} $|T|$ the number of events recorded counting from $1$, thus $\textup{dom}(T)=\{1,\dots,|T|\}$. When $T\colon\mathbb{N}\to\mathbb{R}^d$ is clear from the context, given a dimension $x\leq d$, we use $x\llparenthesis t\rrparenthesis$ as a shorthand for $T(t)(x)$. Given a time interval $(i,j)$ and a \gls{mts} $T$ of at least size $j$, we denote $T[i,\dots,j]$ as its \textit{projection} considering a subsequence of the events occurring within this interval:
$T[i,\dots,j](x)=T(x+i-1) \textup{\,\textbf{if}\,}  x+i-1\in\textup{dom}(T) {\,\textbf{and}\,} 1\leq x\leq j-i+1$.

 Given a  \gls{mts} $T_\mathfrak{E}$ within a training dataset $D$ where  $\textsf{c}$ represents the time-wise classification and a set of maximal and non-overlapping temporal subsequences $[i,j]\in\mathbb{I}_{T_\mathfrak{E}(\textsf{c})}([1,|T|])$ targeting the same event time class value for dimension $\textsf{c}$, we want to learn a function $h$ minimising the classification error $\sum_{i\leq \tau\leq j}|h(T_\mathfrak{E}[i,\dots, j])-T_\mathfrak{E}(\tau)(\textsf{c})|$ for each maximal interval $[i,j]$ and from all $T_\mathfrak{E}$ in $D$. This is a more general formulation than the one posed by current literature, considering time series where $\mathbb{I}_{T_\mathfrak{E}(c)}([1,|T|])=\{[1,|T|]\}$, where one classification label is associated to each  timestamp.

\subsection{Polyadic Logs}\label{sec:pollog}
A \textit{polyadic log} $\mathfrak{S}$ is a pair $\braket{\mathbb{G},\mathcal{L}}$, where \textit{(i)} $\mathcal{L}$ is a collection of distinct \textit{polyadic traces} $\{\sigma^i,\dots,\sigma^n\}$ referring to the auditing of a specific environment $\mathfrak{E}$ of interest (e.g., a \gls{mts}); each trace is a list of temporally ordered \textit{polyadic events} $[\sigma^i_1,\dots,\sigma^i_o]$, where each of these $\sigma^i_j$ is defined as a pair of a set of \textit{durative constituents} and a \textsf{class} $\sigma^i_j=\braket{\{\varsigma^i_{j,1},\dots,\varsigma^i_{j,m}\},\textsf{class}}$, where all constituents start at the same time $j$ but possibly with different activity labels and duration span; in particular, each durative constituent  $\varsigma^i_{j,k}$ is expressed as a triplet  $\braket{\textsf{a},p,\mathcal{s}}$ where $\textsf{a}$  is the activity label, $p$ is  the payload collecting the raw data being associated to the durative constituent, and $\mathcal{s}$ denotes its temporal span s.t. $\mathcal{s}\leq n$. In this, we denote with $\lambda$, $\varpi$, and $\delta$ the function extracting the activity label ($\lambda(\varsigma^i_{j,k})=\textsf{a}$), its payload ($\varpi(\varsigma^i_{j,k})=p$), and its span ($\delta(\varsigma^i_{j,k})=\mathcal{s}$). We denote $\kappa(\varsigma^i_{j,k})$ as the function concatenating all of this information in a finite function $\varpi(\varsigma^\iota_{1,k})\circ[\textup{\_\_label}\mapsto\lambda(\varsigma^\iota_{1,k}),\textup{\_\_span}\mapsto\delta(\varsigma^\iota_{1,k})]$
Last, \textit{(ii)} $\mathbb{G}$ is a collection of \textit{taxonomies} \cite{10.1145/3410566.3410583} represented as Direct (Acyclic) Graphs $G_i = \braket{N_G,R_G,\ell_G}$ rooted in $\ell_G$ determining the relationship between activity labels within the log $\mathfrak{S}$; the entities are represented as a set of nodes $N_G$, and $R_G$ is the set of the is-a relationships. 

\subsection{Poly-DECLARE}

Polyadic traces trivially violate the temporal non-simultaneity assumption from non-polyadic logs \cite{202403.0286} for which each non-polyadic event cannot be associated with multiple activity labels, while now one single polyadic event might contain constituents with different activity labels. As a result, the former algorithms did not consider concurrent violation conditions: e.g., for \textsf{Precedence}(\textsf{A},\textsf{C}), we now also prescribe that a \textsf{C}-labelled constituent shall also never co-occur with an \textsf{A}-labelled constituent  (cfr. the occurrence of $\leq$, Algorithm \ref{algo:polymine}). Achieving all these desiderata requires completely revising the previous algorithms to be adapted to the novel log assumptions (see Algorithm \ref{algo:collectEvidence}). 

Given that each event might contain multiple constituents, we extend the traditional DECLARE semantics by checking whether some or all the constituents activated within a specific event satisfy the given declarative clause. We retain the former as the default interpretation of the declared clauses, thus retaining the same template name while, for indicating the others, we explicitly append a \textsf{All} prefix to the names outlines from \figurename~\ref{tab:dt}. 

\section{\gls{emeritatedf}}
\textit{This section introduces the algorithmic extensions moving from \gls{emeritate} to its \textsc{DataFul (DF)} variant (\gls{emeritatedf}). Unlike our previous contribution, we characterize our algorithms as Explainable and Verified AI. While the A Priori phase adds more contextual information on the data series, the Specification Mining and the White-Box model learning for learning the explainable specifications from the data is carried out in the Ad Hoc phase. The evaluation section provides the Post Hoc evaluation of the trained models under different datasets using the default metrics.}

\subsection{A Priori Explainability}\label{dtmining}
 The a priori explainability phase was improved from \cite{HUO2022117176} as Algorithm~\ref{algo:dtmine}by adequately indexing \gls{mts} to support faster \gls{dt} mining algorithms while also simplifying over the redundant patterns from \figurename~\ref{fig:freighttrain}. Differently from the previous, we extend each constituent having \gls{dt} activity labels, once represented as a dataless durative constituent, with Catch22 features as data payloads for the mined durative constituents.

\subsubsection{Loading and Indexing.}
The first step is to \textit{load} the \gls{mts} discretized as a preliminary pointwise log of events expressing the verification or not of a specific \textit{numerical condition} $x$ expressing different types of numeric variations  (\tablename~\ref{tab:preds}) such as increase (\texttt{i}), absence (\texttt{a}), stationarity (\texttt{s}), and variability (\texttt{v}), and by creating pointwise events satisfying or not some associated condition $P_x(t)$. This will then be used to mine the \gls{dt} patterns.

\begin{table}[!h]
\caption{Point-wise predicates $P_x^\tau$ for discretizing time series $\tau$ over dimension $i$ and representing those as values $\textit{val}_x^\tau$.}\label{tab:preds}
\centering
\begin{tabular}{lrc}
\toprule
$x\quad$& $\textit{val}_x^\tau(t)$ & $P_x^\tau(t)$\\
\midrule
\texttt{i} & $\tau(t+1)(i)-\tau(t)(i)$ & $\textit{val}_\texttt{i}^\tau(t)>0$\\
\texttt{a} & $|\tau(t)(i)|$ & $\textit{val}_\texttt{i}^\tau(t)\leq\varepsilon$\\
\texttt{s} & $|\textit{val}_\texttt{i}(t)|$ & $\textit{val}_\texttt{s}^\tau(t)\leq\varepsilon$\\
\texttt{v} & $\begin{cases}
0 & P_\texttt{s}(t)\\
+\infty & P_\texttt{a}(t)\\
\frac{\textit{val}_\texttt{i}^\tau(t)}{\tau(t)(d)} & \textup{oth.}
\end{cases}$ & $P_\texttt{s}^\tau(t)$\\
\bottomrule
\end{tabular}
\end{table}

We perform a linear scan of each multivariate time series $T_\mathfrak{E}$ so to identify the maximal intervals  $[B,E]\in\mathbb{I}_{T_\mathfrak{E}(\cdot)(\textsf{c})}([1,|T_\mathfrak{E}|])$ associated with the same class reported in dimension \textsf{c} (Line \ref{firstChunk}), thus considering a projection $\tau$ for each of these $[B,E]$. For each dimension of interest (Line \ref{eachDim}), we discretize each dimension and timestamp as in our previous paper \cite{ideas2024a} by considering numerical variations (Line \ref{forxiasv}).  All such constituents are then stored in events (Line \ref{inEvent}) contained in a log $\mathfrak{S}^{i,\tau}_x$ for each dimension $i$, segment $[B,E]$ as $\tau$, and numerical variation $x$ (Line \ref{storeInLog}).

Next, we \textit{index} the previously-discretized logs according to the satisfaction (Line \ref{satChunk}) or violation (Line \ref{violChunk}) of the predicate $P_x^\tau$. We ensure that this information is computed only once by identifying regions of a specific time series' dimension where numerical variation conditions continue to hold or not. As the intervals in such indices are never overlapping, we can store the intervals in order of increasing beginning time and query those in $O(\log|\mathcal{G}^{i,\tau}_{x,\_}|)$ to check whether the time series at dimension $i$ satisfies the numerical variation condition $x$ at a specific time $t$. Hence, we denote and define such query as $t\in \mathcal{G}^{i,\tau}_{x,\_}\Leftrightarrow \exists [b,e]\in \mathcal{G}^{i,\tau}_{x,\_}.\; t\in [b,e]$: this notion is used in the forthcoming subsection.

\begin{algorithm}[!h]
\caption{A Priori Explainability}\label{algo:dtmine}
\begin{spacing}{0.8}
\algrenewcommand\algorithmicindent{.5em}
\begin{algorithmic}[1]
\Procedure{TSLoadIndex}{$T_\mathfrak{E};\varepsilon$}
\ForAll{$[B,E]\in\mathbb{I}_{T_\mathfrak{E}(\cdot)(\textsf{c})}([1,|T_\mathfrak{E}|])$}\label{firstChunk} 
\State $\tau\gets T_\mathfrak{E}[B,E]$;\quad $\overline{\textsf{c}}\gets T_\mathfrak{E}[B,E](B)(\textsf{c})$
\For{$i$ \textbf{from} $1$ to $d-1$}\label{eachDim}
\ForAll{$x\in \{\texttt{i},\texttt{a},\texttt{s},\texttt{v}\}$}\label{forxiasv}
\State $\mathcal{L}^{i,\tau}_x\gets[\sigma^{\mathfrak{E},[B,E],i}_B,\dots,\sigma^{\mathfrak{E},[B,E],i}_E]$ \label{storeInLog}\Comment{\textit{Loading}} 
\ForAll{$1\leq t\leq E-B+1$}
\State $\varsigma^{\mathfrak{E},[B,E],i}_{t,x}\gets\braket{P_x^\tau(t),[\textit{val}\mapsto \textit{val}_x^\tau(t)],1}$ \label{discretize}
\State $\sigma^{\mathfrak{E},[B,E],i}_t\gets\braket{\{\varsigma^{\mathfrak{E},[B,E],i}_{t,x}\},\overline{\textsf{c}}}$\label{inEvent}
\EndFor
\State $\mathcal{G}^{i,\tau}_{x,\textbf{true}}\gets\Set{[b,e]\in \mathbb{I}_{P^\tau_x}([B,E])|\forall b\leq t\leq e. P^\tau_x(t)=\textbf{true}}$\label{satChunk}\Comment{\textit{Indexng}}
\State $\mathcal{G}^{i,\tau}_{x,\textbf{false}}\gets\Set{[b,e]\in \mathbb{I}_{P^\tau_x}([B,E])|\forall b\leq t\leq e. P^\tau_x(t)=\textbf{false}}$\label{violChunk}
\EndFor
\EndFor
\EndFor
\EndProcedure
\algstore{myalg}
\end{algorithmic}
\end{spacing}
\end{algorithm}

\begin{algorithm}[!p]
\begin{spacing}{0.8}
\algrenewcommand\algorithmicindent{.5em}
\begin{algorithmic}[1]
\algrestore{myalg}
\Function{DL}{$\ell,\nu;x,i$}\label{dtlabel}
\State \textbf{if} $\ell=\texttt{S}$ \textbf{then} \Return ($\nu$ \textbf{?} \textsf{IncreaseRapidly} \textbf{:} \textsf{DecreaseRapidly})+\textsf{(dim$_i^x$)}
\State \textbf{if} $\ell=\texttt{HV43}$ \textbf{then} \Return ($\nu$ \textbf{?} \textsf{HighVolatility3} \textbf{:} \textsf{HighVolatility4})+\textsf{(dim$_i^x$)}
\State \textbf{if} $\ell=\texttt{1H}$ \textbf{then} \Return ($\nu$ \textbf{?} \textsf{IncreaseSlowly1} \textbf{:} \textsf{DecreaseSlowly4})+\textsf{(dim$_i^x$)}
\State \textbf{if} $\ell=\texttt{2H}$ \textbf{then} \Return ($\nu$ \textbf{?} \textsf{IncreaseSlowly2} \textbf{:} \textsf{DecreaseSlowly3})+\textsf{(dim$_i^x$)}
\State \textbf{if} $\ell=\texttt{E1}$ \textbf{then} \Return ($\nu$ \textbf{?} \textsf{IncreaseSlowly3} \textbf{:} \textsf{DecreaseSlowly2})+\textsf{(dim$_i^x$)}
\State \textbf{if} $\ell=\texttt{E2}$ \textbf{then} \Return ($\nu$ \textbf{?} \textsf{IncreaseSlowly4} \textbf{:} \textsf{DecreaseSlowly1})+\textsf{(dim$_i^x$)}
\EndFunction
\Statex
\Procedure{DTMineStep}{$\tau, i, x, \nu, [b,e];\mathfrak{E}$}
\State $n\gets e-b+1$
\State  \textbf{yield} $\varsigma^\mathfrak{E}_{b,\texttt{fresh}(b)}:=\braket{\textsc{DL}(\texttt{S},\nu;x,i), \textsc{C22}(\tau^i)\circ \textsc{C22}([b,e]),n}$\label{straightPattern}
\ForAll{$1\leq \mathcal{s}\leq n$}
\ForAll{$b\leq \beta\leq e-\mathcal{s}+1$}
\State $\xi\gets  (e+1)\in\mathcal{G}^{i,\tau}_{x,\textbf{not}\,\nu} \,\textbf{and}\,\beta=e$; $\eta\gets \beta+\mathcal{s}-1$
\If{$\beta=b$}
\If{$(b-1)\in\mathcal{G}^{i,\tau}_{x,\textbf{not}\,\nu}$}\label{line26}
\State \yieldconst{\beta-1}{\eta}{\mathcal{s}+1}{\nu}{\tau^i}{\texttt{1H}}
\If{$\xi$} \State \yieldconst{\beta-1}{\eta+1}{\mathcal{s}+2}{\nu}{\tau^i}{\texttt{HV43}}
\EndIf
\If{$ (\beta-2)\in \mathcal{G}^{i,\tau}_{x,\nu}$}
\State \yieldconst{\beta-2}{\eta}{\mathcal{s}+2}{\nu}{\tau^i}{\texttt{2H}}
\EndIf
\EndIf
\EndIf
\If{$\beta+\mathcal{s}-1=e$ \textbf{and} $\xi$}
\State found $\gets$ \textbf{false}
\If{$[e+1,e+2]\in \mathcal{G}^{i,\tau}_{x,\textbf{not}\,\nu}$ \textbf{and} $e+2\in \mathcal{G}^{i,\tau}_{x,\nu}$}
\State found $\gets$ \textbf{true}
\State \yieldconst{\beta}{\eta+2}{\mathcal{s}+2}{\nu}{\tau^i}{\texttt{E1}}
\EndIf
\If{\textbf{not}\,\textit{found}}
\State \yieldconst{\beta}{\eta+1}{\mathcal{s}+1}{\nu}{\tau^i}{\texttt{E2}}
\EndIf
\EndIf
\EndFor
\EndFor
\EndProcedure
\Statex
\Procedure{DTMine}{$T_\mathfrak{E}$}
\ForAll{$[B,E]\in\mathbb{I}_{T_\mathfrak{E}(\cdot)(\textsf{c})}([1,|T_\mathfrak{E}|])$}
\State $\tau\gets T_\mathfrak{E}[B,E]$
\ForAll{$1\leq j\leq E-B+1$}
\State $\sigma^\mathfrak{E}_{B+j-1,0}:=\braket{\textsf{\_\_raw\_data},[\texttt{dim\_}i\mapsto T_\mathfrak{E}(B+j)(i)]_{1\leq i\leq d-1},1}$\label{rawdata1}
\State $\sigma^\mathfrak{E}\gets\{\sigma^\mathfrak{E}_{B+j-1,0}\}$\label{rawdata2}
\EndFor
\For{$i$ \textbf{from} $1$ to $d-1$}
\State $\tau^i\gets (\tau(1)(i),\dots,\tau(|T_\mathfrak{E}|)(i))$

\ForAll{$x\in \{\texttt{i},\texttt{a},\texttt{s},\texttt{v}\}$}
\ForAll{$[b,e]\in \mathcal{G}^{i,\tau}_{x,\nu}$}

\State $\sigma^\mathfrak{E}\gets\sigma^\mathfrak{E}\cup\,$\Call{DTMineStep}{$\tau,i,x,\textbf{true},[b,e]$}\label{dtmine1}
\State  $\sigma^\mathfrak{E}\gets\sigma^\mathfrak{E}\cup\,$\Call{DTMineStep}{$\tau,i,x,\textbf{false},[b,e]$}\label{dtmine2}
\EndFor
\EndFor
\EndFor
\EndFor
\State \Return $\sigma^\mathfrak{E}$
\EndProcedure

\end{algorithmic}
\end{spacing}
\end{algorithm}

\begin{lemma}[Indexing and Loading Time Complexity] Given a collection of $N$ \gls{mts} of $d$ dimensions with maximum length $t$, the time complexity is in $O(Ndt)$.
\end{lemma}
\begin{proof}
Given that all insertions on indices can be built while linearly scanning the data, and assuming that each $P_x^\tau(x)$ can be tested in constant time, the time complexity of scanning each maximal time interval $[B,E]$ and each value in it boils down to the maximal time series length $t$. Given that the number of point-wise predicates is constant, we can conclude that the linear scan is provided for each dimension and \gls{mts} within the collection, thus leading to $O(Ndt)$. \qed
\end{proof}

\subsubsection{Indexed \gls{dt} Mining.} This phase enables the representation of a time series as a polyadic trace, where the first constituent for each $j$-th event reflects the values associated with each dimension at time $j$, i.e. $T_\mathfrak{E}(j)$ (Lines \ref{rawdata1}-\ref{rawdata2}), while the remainder is derived from the \gls{dt}-mined pointwise constituents from the previous phase. The composition of the $\mathcal{G}$-indexed intervals generate \gls{dt} patterns distinguishable from a constituent activity label (Line \ref{dtlabel}), where the name of the \gls{dt} pattern reports the dimension $i$ and the specific numerical variation $x$. The algorithm proceeds by scanning each maximal interval $[B,E]$ for a time series $T_\mathfrak{E}$ over each dimension $i$, numerical variation $x$, and $\mathcal{G}$-indexed satisfiability interval $[b,e]$ (Lines \ref{dtmine1}-\ref{dtmine2}). The segmentation of the time series through the $\mathcal{G}$ indices allows to promptly derive that any subsequent or preceding interval within the same dimension will violate some $x$ constraints if the current interval satisfies them, and vice versa. Differently from our previous paper, for each mined constituent referring to a specific dimension $i$ and interval $[\beta,\eta]\subseteq [b,e]$, we generate a payload containing the Catch22 \cite{catch24}  features (\textsf{C22}) capturing the dynamic properties of the $i$-th dimension in $\tau^i[\beta,\eta]$, which are paired with the timestamp fluctuations. By doing so, we generalise over the CIF classifier \cite{9378424}, which focuses on distinguishing time series from the behaviours occurring within a specific fixed-size sliding window while potentially considering the features for all possible sliding windows captured as \gls{dt} patterns. The computation of the Catch24-payload is parallelized due to the high computational nature of the metrics used to describe the numerical fluctuations of the data. The main mining algorithm considers a subset of all the \glspl{dt} patterns described in Section \ref{tsdec}, where some high volatility patterns are discarded so to favour the recognition of shorter volatility patterns rather than preferring joining prolonged variations. Overall, This helps to reduce the number of patterns to be returned through the combination of subsequent patterns while reducing the time to mine the clauses without considerably undermining the predictive power of the classifier. At the end of the process, each time series $T^\mathfrak{E}$ is discretized into a single polyadic trace $\sigma^\mathfrak{E}$.

\begin{lemma}
    The worst-case time complexity for the indexed DT mining is superpolynomial over the maximum length of the segmented interval $\mathcal{s}$ and linear over the number and size of the \gls{mts}. 
\end{lemma}
\begin{proof}
    Note that the worst-case scenario time complexity does not happen for \gls{mts} exhibiting stable values, as this would lead to just one maximal interval $[B,E]$ per class. In fact, these situations have $\xi$  and the condition at Line \ref{line26}
 as false, as there will be no other neighbouring intervals to consider. This will only lead to returning the durative constituent at Line \ref{straightPattern}, where still Catch22 statistics are computed. As the time complexity of computing all the Catch22 statistics depends on the length of the interval of interest, we denote this as $C_\mathcal{s}$.

 Thus, the worst time complexity happens when we have at least two intervals, leading not only to the generation of the aforementioned durative constituent, but potentially to other patterns occurring within the nested iteration. Given that within the nested loop leading to $O(\mathcal{s}^2)$ iterations we test queries having a time in $O(\log\mathcal{s})$, we obtain an overall time complexity of $O(\mathcal{s}^2\log\mathcal{s}C_\mathcal{s}\frac{t}{\mathcal{s}}N)$. \qed
 \end{proof}

\subsubsection{Serialization} At serialization time, we avoid writing any potential redundant constituent having the same activation label but referring to a shorter time interval to the other sibling constituents, thus capturing the behaviour associated with the longest span. We generate a collection of taxonomies for reconducing each constituent label to the original \gls{mts} dimension the pattern was referring to: we generate a taxonomy $G_i=\braket{N_G,R_G,\ell_G}$ for each dimension $i$, where $N_G$ contains all the \glspl{dt} associated with $\textit{dim}^x_i$ thus including $\textit{dim}^x_i$ acting as the root $\ell_G$; $R_G$ connects only the root with the associated \gls{dt} labels. Given the impossibility of serializing such data in XES \cite{XES} due to their strict assumptions on the log representation, we serialize the resulting information on a customary representation in \texttt{json}.

\subsection{Ad Hoc Explainability}\label{sec:polymine}

\textit{We load the single log composed of many traces with classes changing in time as many segmented logs $\mathfrak{S}_y$ as the total number of classes distinctly occurring by segmenting the traces in maximal intervals where the same class holds. Then,  differently from our previous solution, where each segmented log was mined separately via Bolt2 \cite{bolt2} to then derive the trace embedding in a later phase, we contemporarily load and mine all the segmented logs containing maximal sequences referring to the same classes. We then proceed to the DECLARE clause refinement for the ones shared across the logs as a result of the mining phase: as generic dataless DECLARE clauses with \gls{dt} activity labels might be not sufficient to temporally characterize the classes, we try to specialize each clause according to the specificity of each log/class. The rationale is the following: if the same dataless clause is frequently mined from two segmented logs, at least one trace exists from each of these where the dataless clause is both activated and holds. Then, the only way to differentiate the clause across the two segmented logs is to change the satisfiability outcome of the trace by refining the original dataless clause with dataful predicates for either the activation or the target condition. Suppose the payload referring to the constituents activated by the clause can be differentiated. In that case, we can refine the original dataless clause as two dataful ones, each with an activation condition ($p$ and $p'$) satisfying only the activations from one segmented log. This leads to one dataful clause being then activated over the traces of one segmented log, but being vacuously satisfied by the traces of the other. If that will not suffice, then the only further possible way to differentiate is to generate target predicates similar to those above: in this other scenario, the missed satisfaction of the target condition will lead to a violation of the clause, thus further refining the trace behaviour in terms of clause (un)satisfaction.}

\textit{This significantly differs from other deviance mining approaches \cite{LENO2020101482}, where activation conditions are mainly clustered not necessarily according to the target classes of interest \cite{Undone}, thus not achieving a truthful clause separation for differentiating them across the distinct logs as discussed above. By exploiting a decision tree as a classifier, we can ensure that the data predicates used to differentiate the classes will summarize the data properties as unary dataful predicates, which might reconstructed by traversing each branch of the learned decision tree by interpreting it as a single propositional formula $\pi$.}

\subsubsection{Trace Loading and Segmentation.}\label{partandload}
Similarly to our previous paper, for each trace $\sigma^{\mathfrak{E}_A}$  in the \texttt{json}-serialized log $\mathfrak{S}$ describing a distinct \gls{mts}, we scan all polyadic events, and we group them by maximal contiguous sequences associated to the same class $y$. Each class $y$ will then correspond to one single final log $\mathfrak{S}_y$, where each maximal contiguous sequence of polyadic events will correspond to one single new polyadic trace in $\mathfrak{S}_y$. As this transformation also preserves the \textsf{\_\_raw\_data} information associated with each polyadic event, we can then later on collect just the raw data associated with the polyadic traces for later on running atemporal event classification tasks by just correlating the class within the \textsf{\_\_raw\_data} payload with the other available values. 

\begin{algorithm}[!h]
\caption{
Ad Hoc Explainability}\label{algo:dataawaremine}
\begin{spacing}{0.8}
\algrenewcommand\algorithmicindent{.5em}
\begin{algorithmic}[1]
\renewcommand\theContinuedFloat{\alph{ContinuedFloat}}
\Function{PolyadicDataAware}{$\mathfrak{S}_1,\dots,\mathfrak{S}_n;\theta$}
\State $\Sigma\gets \bigcup_{\mathfrak{S}_\iota=\braket{\mathbb{G},\mathcal{L}},\braket{N_G.R_G,\ell_G}\in\mathbb{G}}N_G$ \Comment{\textit{Determining the alphabet}}
\State $\Phi\gets\emptyset$ \Comment{\textit{Clauses not requiring dataful refinement}}
\ForAll{$1\leq \iota\leq n$}
\ForAll{$\sigma^i\in \mathcal{L},\mathfrak{S}_\iota=\braket{\mathbb{G},\mathcal{L}}$}
\State DataFrame[${(\iota,i),\texttt{clazz}}$]:=$\iota$\Comment{\textit{Segmented trace/class assoc.}}\label{clazz}
\EndFor
\State $P_\iota\gets $\Call{FrequentItemsets}{$\mathfrak{S}_\iota,\theta$} \Comment{\textit{Maximal length of 2}}\label{fiset}
\State $\textsf{FreqPairs}_\iota,\Phi_\iota\gets$\Call{GenerateUnaryClauses}{$P_\iota$} \Comment{\textit{Unary clauses} \cite{bolt2}}\label{genUnary}
\EndFor
\State $\Phi\gets$ \Call{UnaryRefine}{$\Phi_1,\dots,\Phi_n,\Phi;\mathfrak{S}_1,\dots,\mathfrak{S}_n$} \Comment{Algorithm \ref{algo:unaryrefinement}}
\Statex 
\State FP$\gets\bigcup_{1\leq\iota\leq n}$FreqPairs$_\iota$
\ForAll{$\braket{A,B}\in$FP}\label{allPairs}
\State $C\gets \{\star(A,B)|\star\in\textit{BinaryClauses}\}$\Comment{\textit{Maximising recall by probing all clauses}}
\If{$\exists!j. \braket{A,B}\in$FreqPairs$_j$}\label{norefine}
\State $\Phi\gets\Phi\cup C$
\Else
\State $\Phi\gets\Phi\cup $\,\Call{BinaryRefine}{${A},{B},\mathfrak{S}_1,\dots,\mathfrak{S}_n;\textbf{true}$}\Comment{Algorithm \ref{algo:polymine}}\label{attemptPolyRefine}
\EndIf
\EndFor 
\Statex

\State \textbf{global} DataFrame
\ForAll{$1\leq \iota\leq n$}\Comment{\textit{Dataless \cite{ideas2024a}}}
\ForAll{$\sigma^i\in \mathcal{L},\mathfrak{S}_\iota=\braket{\mathbb{G},\mathcal{L}}$}
\ForAll{$\textsf{clause}\in\Phi$}
\State DataFrame[${(\iota,i),\textsf{clause}}$]:=$\mathbf{1}_{	\sigma^i\models\textsf{clause}}-\mathbf{1}_{\sigma^i\not\models\textsf{clause}}$ 	\label{dcs}
\EndFor
\EndFor
\EndFor
\State \Return \Call{DecisionTree}{$\{([k\mapsto v]_{},\textit{row}[\texttt{clazz}])|\textit{row}\in \textrm{DataFrame}\}$}
\EndFunction
\end{algorithmic}
\end{spacing}
\end{algorithm}

\subsubsection{Polyadic Deviant Model Learning.} Algorithm \ref{algo:dataawaremine} showcases the procedure for extracting an explainable specification capturing \gls{mts} classes' temporal characterization (Line \ref{clazz}). This is achieved by considering DECLARE clauses, be they data or dataless, as features for describing class-segmented traces. By interpreting the numerical values as a violation, satisfaction, or vacuous satisfaction of a specific clause, we then use a white box classifier such as Decision Tree to extract a propositional representation for the clauses' satisfaction.

Similarly to our previous implementation \cite{bolt2,ideas2024a}, we consider all the most frequent itemsets for at most two activity labels and support of at least $\theta$ (Line \ref{fiset}). Unary DECLARE clauses such as \textsf{Init}, \textsf{End}, and \textsf{Exists} are mined from the frequently-occurring unary patterns as per Bolt2 (Line \ref{genUnary}). We further proceed on the refinement for each of the clauses into their dataful counterpart if and only if the payload-based refinement allows us to substantially differ the traces' behaviour according to the constituents' payloads and, otherwise, we backtrack to the original dataless variants (Line \ref{dcs}). Similar considerations can be carried out for the remaining binary clauses: given all the possible binary frequent patterns occurring across all logs (Line \ref{allPairs}), we perform no refinement if such pair is frequent only in one segmented log (Line \ref{norefine}) and, otherwise, we proceed by refining the clauses when possible (Line \ref{attemptPolyRefine}). Last, we train the white-box classifier over the extracted embedding for each class-segmented trace, from which we derive a propositional and declare-based characterization for each \gls{mts} class.

\begin{algorithm}[!p]
\caption{Ad Hoc Explainability: unary clause dataful refinement}\label{algo:unaryrefinement}

\begin{spacing}{0.8}
\algrenewcommand\algorithmicindent{.5em}
\begin{algorithmic}[1]
            \ContinuedFloat 
\Function{RefineAttempt}{$D,\mathcal{D},L,n;\vartheta=.7$}
\State \textbf{global} DataFrame
\State $D_\textrm{train},D_\textrm{test}\gets$\Call{Split}{$D,\vartheta$}
\State $\mathcal{M}\gets \Call{DecisionTree}{D_\textrm{train}}$
\State $\mathcal{Y}\gets [\iota\mapsto \bigwedge_{\pi\in\mathcal{M},\pi=H\Rightarrow \iota}H]_{1\leq\iota\leq n}$
\State \algorithmicif\, \Call{Accuracy}{$\mathcal{M},D_\textrm{test}$}$\leq50\%$ \algorithmicthen\, \Return \textbf{false} \label{discard} 
\ForAll{$(\iota,\sigma^i)\in\dom(\mathcal{D})$}
\ForAll{$\pi\in\mathcal{M}$}\label{eachPath}
\State DataFrame[${(\iota,i),\textsf{All}L(\star,\pi)}$]:=$\mathbb{B}(|\mathcal{D}(\iota,\sigma^i)|=\sum_{p\in \mathcal{D}(\iota,\sigma^i)}\mathbf{1}_{\pi(p)})$ \label{allConstit}
\State DataFrame[${(\iota,i),L(\star,\pi)}$]:=$\mathbb{B}(0<\sum_{p\in \mathcal{D}(\iota,\sigma^i)}\mathbf{1}_{\pi(p)})$ \label{justSome}
\EndFor
\ForAll{$\iota\in\dom(\mathcal{Y})$}\label{remainingLoop}
\State $\pi\gets \mathcal{Y}(\iota)$
\State DataFrame[${(\iota,i),\textsf{All}L(\star,\pi)}$]:=$\mathbb{B}(|\mathcal{D}(\iota,\sigma^i)|=\sum_{p\in \mathcal{D}(\iota,\sigma^i)}\mathbf{1}_{\pi(p)})$ 
\State DataFrame[${(\iota,i),L(\star,\pi)}$]:=$\mathbb{B}(0<\sum_{p\in \mathcal{D}(\iota,\sigma^i)}\mathbf{1}_{\pi(p)})$ 
\EndFor
\EndFor
\State \Return \textbf{true}
\EndFunction
\Statex
\Function{UnaryRefine}{$\Phi_1,\dots,\Phi_n,\Phi;\mathfrak{S}_1,\dots,\mathfrak{S}_n$}
\State refineInit $\gets \exists\textsf{a}\in\Sigma.\exists i,j\in\mathbb{N}. \textit{Init}(\textsf{a},	\textbf{true})\in\Phi_i\cup\Phi_j $	\label{refineInit}
\If{refineInit}\Comment{\textsf{Init}\textit{ refinement} }
\State $\mathcal{D}\gets	\left[(\iota,\sigma^i)\mapsto S\right]_{1\leq\iota\leq n,\mathfrak{S}_\iota=\braket{\mathbb{G},\mathcal{L}},\sigma^i\in\mathcal{L},\sigma^i_1=\braket{S,\iota}}$
\State $D\gets\Set{(\kappa(\varsigma^i_{1,k}),\iota)|\varsigma^i_{1,k}\in \sigma^i_1,\sigma^i\in \mathcal{L},\mathfrak{S}_\iota=\braket{\mathbb{G},\mathcal{L}},1\leq\iota\leq n}$\label{collBegin}
\State refineInit $\gets$ \Call{RefineAttempt}{$D,\mathcal{D},\textsf{Init},n$}
\EndIf
\State \algorithmicif\, \textbf{not} refineInit \algorithmicthen\, $\Phi\gets\Phi\cup\{\textit{Init}(\textsf{b},\textbf{true})|\textit{Init}(\textsf{b},\textbf{true})\in \bigcup_{1\leq\iota\leq n}\Phi_\iota\}$
\Statex
\State refineEnd $\gets \exists\textsf{a}\in\Sigma.\exists i,j\in\mathbb{N}. \textit{End}(\textsf{a},	\textbf{true})\in\Phi_i\cup\Phi_j $\label{refineEnd}
\If{refineEnd}\Comment{\textsf{End} \textit{ refinement} }
\State $\mathcal{D}\gets	\left[(\iota,\sigma^i)\mapsto S\right]_{1\leq\iota\leq n,\mathfrak{S}_\iota=\braket{\mathbb{G},\mathcal{L}},\sigma^i\in\mathcal{L},\sigma^i_{|\sigma^i|}=\braket{S,\iota}}$
\State  $D\gets\{(\kappa(\varsigma^i_{|\sigma^i|,k}),\iota)\;|\;\varsigma^i_{|\sigma^i|,k}\in \sigma^i_{|\sigma^i|},\sigma^i\in \mathcal{L},\mathfrak{S}_\iota=\braket{\mathbb{G},\mathcal{L}},1\leq\iota\leq n\}$\label{collEnd}
\State refineEnd $\gets$ \Call{RefineAttempt}{$D,\mathcal{D},\textsf{End},n$}
\EndIf
\State \algorithmicif\, \textbf{not} refineEnd \algorithmicthen\, $\Phi\gets\Phi\cup\{\textit{End}(\textsf{b},\textbf{true})|\textit{End}(\textsf{b},\textbf{true})\in \bigcup_{1\leq\iota\leq n}\Phi_\iota\}$
\Statex
\State refineEx $\gets \exists\textsf{a}\in\Sigma.\exists i,j\in\mathbb{N}. \textit{Exists}(\textsf{a},	\textbf{true})\in\Phi_i\cup\Phi_j $\label{refineExists}
\If{refineEx} \Comment{\textsf{Exists} \textit{ refinement} }
\State $\mathcal{D}\gets	\left[(\iota,\sigma^i)\mapsto\Set{\varpi(\varsigma^i_{j,\_})|\varsigma^i_{j,\_}\in \sigma^i_{j},\sigma^i\in \mathcal{L}}\right]_{1\leq\iota\leq n,\mathfrak{S}_\iota=\braket{\mathbb{G},\mathcal{L}}}$
\State $D\gets\Set{(\kappa(\varsigma^i_{j,k}),\iota)|\varsigma^i_{j,k}\in \sigma^i_{j},\sigma^i\in \mathcal{L},\mathfrak{S}_\iota=\braket{\mathbb{G},\mathcal{L}},1\leq\iota\leq n}$\label{collExists}
\State refineEx $\gets$ \Call{RefineAttempt}{$D,\mathcal{D},\textsf{Exists},n$}
\EndIf
\State \algorithmicif\, \textbf{not} refineEx \algorithmicthen\, $\Phi\gets\Phi\cup\{\textit{Exists}(\textsf{b},\textbf{true})|\textit{Exists}(\textsf{b},\textbf{true})\in \bigcup_{1\leq\iota\leq n}\Phi_\iota\}$
\State \Return $\Phi$
\EndFunction
\end{algorithmic}
\end{spacing}
\end{algorithm}

\paragraph*{Unary Refinement.} Algorithm \ref{algo:unaryrefinement} outlines the dataful refinement of the unary clauses of \textsf{(All)Init}, \textsf{(All)End}, and \textsf{(All)Exists}, which is attempted if and only if we can mine the same clause from at least two class-segmented logs (Lines \ref{refineInit}, \ref{refineEnd}, and \ref{refineExists}). When this occurs, we collect all the payloads associated with the constituents occurring at the beginning (Line \ref{collBegin}), end (Line \ref{collEnd}) or from any constituent (Line \ref{collExists}) according to the clause to be refined. Lines \ref{allConstit} and \ref{justSome} achieve the polyadic extension for DECLARE by considering the \textsf{All$\star$} clause variaties. We also consider all the predicates given within the decision tree and representing one class of interest and put them in disjunction, thus considering conditions generally identifying classes (Line \ref{remainingLoop}). No refinement is provided if the decision tree cannot adequately separate the activation payloads according to the associated class to generate refined dataful activation predicates  (Line \ref{discard}). 

As the explanation generated from the decision tree's \textit{ex post} phase provides a propositional formula of declarative clauses, we discard the refinement for \textsf{Absence}, as this can still appear as the negation of any \textsf{Exist} clause.

\begin{algorithm}[!p]
\caption{Ad Hoc Explainability: binary clause dataful refinement}\label{algo:polymine}
\begin{spacing}{0.8}
\algrenewcommand\algorithmicindent{.5em}
\begin{algorithmic}[1]
\Procedure{FillInDataFrame}{$S^\iota_i,\ell$}
\State \textbf{global} DataFrame
\If{$S^\iota_i=\emptyset$ \textbf{or} $S^\iota_i={\textsf{Vac}}$}
\State DataFrame[$(\iota,i),\ell$]:=0\label{deactivate}
\ElsIf{$\textsf{Viol}\in S^\iota_i$}
\State DataFrame[$(\iota,i),\ell$]:=-1\label{aviol}
\Else
\State DataFrame[$(\iota,i),\ell$]:=1\label{issat}
\EndIf
\EndProcedure
\Statex

\Function{BinaryRefine}{$\textsf{A},\textsf{B},\mathfrak{S}_1,\dots,\mathfrak{S}_n;\texttt{poly}=\textbf{true}$}
\State template $\gets$ $[\textsf{ChainResponse},\textsf{ChainPrecedence}, \textsf{Precedence},\textsf{Response}]$
\State shorthands $\gets$ $[\textsf{cr},\textsf{cp},\textsf{p},\textsf{r}]$
\State dictionary $\gets$\Call{zip}{template,shorthands}
\State pairs $\gets\{\braket{\textsf{A},\textsf{B}},\braket{\textsf{B},\textsf{A}}\}$
\ForAll{$1\leq\iota\leq n$}
\State $\mathfrak{S}_\iota=\braket{\mathbb{G},\mathcal{L}_\iota}$
\State {\color{blue}\textbf{poly} $\gets $\algorithmicif\;(\texttt{poly}\;\textbf{and}\; $\exists G\in\mathbb{G}. \mathsf{A},\mathsf{B}\in N_G$)\;\algorithmicthen\;\textbf{true}\;\algorithmicelse\;\textbf{false}}\label{consideringPoly}
\State  \Call{Chains}{\textsf{A}$'$,\textsf{B}$'$,$\theta$,\texttt{poly},\texttt{heur};$\mathfrak{S}_y$}  \Comment{Algorithm \ref{algo:collectEvidence}} \State \Call{RespPrec}{\textsf{A}$'$,\textsf{B}$'$,$\theta$,\texttt{poly},\texttt{heur};$\mathfrak{S}_y$}\label{reAB}  \Comment{Algorithm \ref{algo:collectEvidence}}

\EndFor
\ForAll{$\textsf{c}(\textsf{A}',\textsf{B}')$ \textsf{s.t.} $\braket{\textsf{A}',\textsf{B}'}\in${pairs} \textbf{and} $c\in$template}
\State short $\gets$ dictionary[\textsf{c}]
\State $D_\textup{act}\gets\Set{(\kappa(\varsigma^i_{j,k}),\iota)|\varsigma^i_{j,k},\braket{\varsigma^i_{j,k},\_}\in\textup{act}(\iota)^\textup{
short}_{\textsf{A}',\textsf{B}'},1\leq \iota\leq n}$\label{actCollect}
\If{$D_\textup{act}=\emptyset$} \textbf{yield} $\textsf{c}(\textsf{A}',\textsf{B}')$
\Else 
\State $\mathcal{M}\gets$\Call{DecisionTree}{$D_\textup{act}$} 
\If{\Call{Purity}{$\mathcal{M}$}$>50\%$}\label{pure1}
\ForAll{$\pi\in\mathcal{M}$}\Comment{\textit{Refine by activations}}
\ForAll{$1\leq\iota\leq n,\;\mathfrak{S}_\iota=\braket{\_,\mathcal{L}},\;\varsigma^i_{j,k},\braket{\varsigma^i_{j,k},\_}\in\textup{act}(\iota)^\textup{
short}_{\textsf{A}',\textsf{B}'}$}
\State test$\gets\pi(\kappa(\varsigma^i_{j,k}))$\label{beginTest}
\If{$ \varsigma^i_{j,k}\in \textup{viol}(\iota)^\textup{
short}_{\textsf{A}',\textsf{B}'}$}
\State $S_i^\iota\gets S_i^\iota\cup\{\textup{test}\,\textbf{?}\,\textsf{Viol}\,\textbf{:}\,\textsf{Vac}\}$
\Else
\State $S_i^\iota\gets S_i^\iota\cup\{\textup{test}\,\textbf{?}\,\textsf{Sat}\,\textbf{:}\,\textsf{Vac}\}$\label{endTest}
\EndIf
\EndFor
\State \algorithmicforall\, $\sigma^i\in\mathfrak{S}_\iota$ \algorithmicdo\, \Call{FillInDataFrame}{$S^\iota_i, \textsf{c}(\textsf{A}',\pi,\textsf{B}',\textbf{true})$}

\EndFor
\Else\, 
\State $D_\textup{tgt}\gets\Set{(\kappa(\varsigma^i_{h,k'}),\iota)|\braket{\varsigma^i_{j,k},\varsigma^i_{h,k'}}\in\textup{act}(\iota)^\textup{
short}_{\textsf{A}',\textsf{B}'},1\leq \iota\leq n}$\label{tgtCollect}

\State $\mathcal{M}\gets$\Call{DecisionTree}{$D_\textup{tgt}$}

\If{\Call{Purity}{$\mathcal{M}$}$>50\%$}\label{pure2}
\ForAll{$\pi\in\mathcal{M}$}\Comment{\textit{Refine by targets}}
\ForAll{$1\leq\iota\leq n,\;\mathfrak{S}_\iota=\braket{\_,\mathcal{L}},\;\varsigma^i_{j,k},\braket{\varsigma^i_{j,k},\_}\in\textup{act}(\iota)^\textup{
short}_{\textsf{A}',\textsf{B}'}$}
\State test$\gets\pi(\kappa(\varsigma^i_{j,k}))$
\If{$ \varsigma^i_{j,k}\in \textup{viol}(\iota)^\textup{
short}_{\textsf{A}',\textsf{B}'}$}
\State $S_i^\iota\gets S_i^\iota\cup\{\textup{test}\,\textbf{?}\,\textsf{Viol}\,\textbf{:}\,\textsf{Vac}\}$
\Else
\State $S_i^\iota\gets S_i^\iota\cup\{\textup{test}\,\textbf{?}\,\textsf{Sat}\,\textbf{:}\,\textsf{Vac}\}$
\EndIf
\EndFor
\State \algorithmicforall\, $\sigma^i\in\mathfrak{S}_\iota$ \algorithmicdo\, \Call{FillInDataFrame}{$S^\iota_i, \textsf{c}(\textsf{A}',\textbf{true},\textsf{B}',\pi)$}\label{backtracking}

\EndFor
\Else\, \textbf{yield} $\textsf{c}(\textsf{A}',\textsf{B}')$  \Comment{\textit{Backtracking to the dataless clause}}
\EndIf
\EndIf
\EndIf
\EndFor


%

\EndFunction

\end{algorithmic}
\end{spacing}
\end{algorithm}

\paragraph*{Binary Refinement.} Algorithm \ref{algo:polymine} describes refining some Polyadic DECLARE binary clauses of choice. Given that the clauses expressing \textsf{(Excl)Choice}, \textsf{CoExistence}, and \textsf{RespExistence} can be easily formulated as propositional formulas composing \textsf{Exists}, we focus on refining the clauses that cannot be characterized through the process above: \textsf{Precedence}, \textsf{Response}, \textsf{ChainPrecedence}, and \textsf{ChainResponse}. We also discard \textsf{ChainSuccession} (and \textsf{Succession}) as those are composite clauses derivable from the former by conjunction, thus being also derivable from a propositional formula. For the time being, we do  not consider \textsf{All}$\star$ template variants, thus considering clause satisfaction if there are no violations across all activations within a trace.

After recalling that this procedure is invoked for any pair of activity labels \textsf{A} and \textsf{B}  appearing as frequent in at least two segmented logs (Algorithm \ref{algo:dataawaremine}, Line \ref{attemptPolyRefine}), we first collect the evidence for the fulfilment of the aforementioned dataless clauses (narrated in the next paragraph). As this enables the collection of all the activated (Line \ref{actCollect}) and targeted (Line \ref{tgtCollect}) constituents' payloads, we can then see if is possible to extract a propositional characterization of the data through a decision tree; we consider the data summarization process through propositionalization successful if we tree achieves a suitable amount of purity (Lines \ref{pure1} and \ref{pure2}). After extracting each path in such a tree as a binary predicate $\pi$, we consider all the activations first within a single trace $\sigma^i$ of a segmented log $\mathfrak{S}_\iota$: differently from the refinement phases performed in DML \cite{Undone} and in event-based mining algorithms \cite{LENO2020101482}, we change the clause satisfaction according to the joint fulfilment of the predicate $\pi$ to generate a correct representation of the embedding: activated conditions that are no more satisfying $\pi$ will always lead to a non-activation of the clause, and otherwise we retain the violation/satisfaction condition (Lines \ref{beginTest}-\ref{endTest}). We then consider the trace as violating the refined clause if at least one constituent leads to its violation (Line \ref{beginTest}); we consider the clause vacuously satisfied if was never activated before or if the addition of the new $\pi$ activation condition leads to de-activation of the clause testing (Line \ref{deactivate}), and we consider as the trace being globally satisfied by the clause otherwise (Line \ref{issat}). We draw similar considerations for the refinement of the target conditions. If neither of these two attempts at refining is satisfactory, we backtrack to the dataless clause (Line \ref{backtracking}). 

\begin{algorithm}[!p]
\caption{Polyadic Mining for (Chain)Response and (Chain)Precedence.}\label{algo:collectEvidence}
\begin{spacing}{0.8}
\algrenewcommand\algorithmicindent{.5em}
\begin{algorithmic}[1]

\Procedure{Chains}{$\mathsf{A},\mathsf{B},\theta,\texttt{poly};\mathcal{L}_y$}

\ForAll{$\sigma^i\in\mathcal{L}_y$}
\ForAll{$\varsigma^i_{j,k}$ \textbf{s.t.} $\lambda(\varsigma^i_{j,k})=\mathsf{A}$}
\State  {\color{blue}$\textsf{span}\gets $\algorithmicif\;\texttt{poly}\;\algorithmicthen\; $\pi(\varsigma^i_{j,k})$\;\algorithmicelse\;$1$ }\label{ignoreSpan1}
\State \algorithmicif\; $\not\exists k.\lambda(\varsigma^i_{j\color{blue}{\boldsymbol{+}\textsf{span}},k})=\textsf{B}$ \algorithmicthen\; \{\ ${\textup{act}(y)}^{\textsf{cr}}_{\mathsf{A},\mathsf{B}}.\textup{add}(\varsigma^i_{j,k})$; $\textup{viol}(y)^{\textsf{cr}}_{\mathsf{A},\mathsf{B}}.\textrm{add}(\varsigma^i_{j,k})$\} 
\State \algorithmicelse\, ${\textup{act}(y)}^{\textsf{cr}}_{\mathsf{A},\mathsf{B}}.\textup{add}(\braket{\varsigma^i_{j,k},\varsigma^i_{j\color{blue}{\boldsymbol{+}\textsf{span}},k}})$ \textbf{for all} $k$ \textbf{s.t.} $\lambda(\varsigma^i_{j\color{blue}{\boldsymbol{+}\textsf{span}},k})=\textsf{B}$.
\If{$j>1$}
\State  ${\textup{act}(y)}^{\textsf{cp}}_{\mathsf{A},\mathsf{B}}.\textup{add}(\varsigma^i_{j,k})$; 
\If{$\not\exists h,k'.\lambda(\varsigma^i_{h,k'})=\textsf{B}$ \;\textbf{and}\; $h+{\color{blue}(\texttt{poly}\textbf{\,\textrm{?}\,}\delta(\varsigma^i_{h,k'})\textrm{\textbf{\,:\,}}1)}{\color{blue}\leq} j$}
\State  $\textup{viol}(y)^{\textsf{cp}}_{\mathsf{A},\mathsf{B}}.\textrm{add}(\varsigma^i_{j,k})$
\Else \,   ${\textup{act}(y)}^{\textsf{cr}}_{\mathsf{A},\mathsf{B}}.\textup{add}(\braket{\varsigma^i_{j,k},\varsigma^i_{h,k'}})$ \textbf{for all} $h,k'$ \textbf{s.t.} $\lambda(\varsigma^i_{h,k'})=\textsf{B}$ \;\textbf{and}\; $h+{\color{blue}(\texttt{poly}\textbf{\,\textrm{?}\,}\delta(\varsigma^i_{h,k'})\textrm{\textbf{\,:\,}}1)}{\color{blue}\leq} j$
\EndIf

\EndIf
\EndFor
\EndFor
\EndProcedure
\Statex
\Procedure{RespPrec}{$\mathsf{A},\mathsf{B},\theta,\texttt{poly},\texttt{heur};{\mathcal{L}_y}$}
\If{$\not\exists \varsigma^i_{h,k}. \lambda(\varsigma^i_{h,k})=\textsf{A}$}
\State \Return \Comment{\textit{Not reporting  vacuous satisfaction explicitly}}
\ElsIf{$\not\exists \varsigma^i_{h,k}. \lambda(\varsigma^i_{h,k})=\textsf{B}$}
\ForAll{$\varsigma^i_{j,k}$ \textsf{s.t.} $\lambda(\varsigma^i_{j,k})=\textsf{A}$}
\State ${\textup{act}(y)}^\textsf{r}_{\mathsf{A},\mathsf{B}}.\textup{add}(\varsigma^i_{j,k})$; $\textup{viol}(y)^\textsf{r}_{\mathsf{A},\mathsf{B}}.\textup{add}(\varsigma^i_{j,k})$; ${\textup{act}(y)}^\textsf{p}_{\mathsf{A},\mathsf{B}}.\textup{add}(\varsigma^i_{j,k})$;
\EndFor
\Else
\ForAll{$\sigma^i\in\mathcal{L}_y$ \textsf{s.t.} $\exists \varsigma^i_{j,k}.\; \lambda(\varsigma^i_{j,k})=\textsf{A}\vee \lambda(\varsigma^i_{j,k})=\textsf{B}$}
\If{$\not\exists {h,k'}. \:\lambda(\varsigma^i_{h,k'})=\textsf{B}$}\Comment{\textit{Only }\textsf{A}\textit{s}}
\State  ${\textup{act}(y)}^\textsf{r}_{\mathsf{A},\mathsf{B}}.\textup{add}(\varsigma^i_{j,k})$; $\textup{viol}(y)^\textsf{r}_{\mathsf{A},\mathsf{B}}.\textup{add}(\varsigma^i_{j,k})$;
\State  ${\textup{act}(y)}^\textsf{p}_{\mathsf{A},\mathsf{B}}.\textup{add}(\varsigma^i_{j,k})$; $\textup{viol}(y)^\textsf{p}_{\mathsf{B},\mathsf{A}}.\textup{add}(\varsigma^i_{j,k})$;
\ElsIf{$\not\exists {h,k'}. \:\lambda(\varsigma^i_{h,k'})=\textsf{A}$}\Comment{\textit{Only }\textsf{B}\textit{s}}
\State  ${\textup{act}(y)}^\textsf{p}_{\mathsf{B},\mathsf{A}}.\textup{add}(\varsigma^i_{j,k})$;  $\textup{viol}(y)^\textsf{p}_{\mathsf{A},\mathsf{B}}.\textup{add}(\braket{\texttt{NULL}, \varsigma^i_{j,k}})$
\ElsIf{$\lambda(\varsigma^i_{j,k})=\textsf{A}$}\Comment{\textit{Both occur}}
 \State ${\textup{act}(y)}^\textsf{p}_{\mathsf{A},\mathsf{B}}.\textup{add}(\sigma^i_{j,k})$; 
\If{$\exists \varsigma^i_{j+h,k},\varsigma^i_{j,k'}.\; h{\color{blue}\geq 0} \textbf{\,and\,} \lambda(\varsigma^i_{j+h,k})=\textsf{B} \textbf{\,and\,} \lambda(\varsigma^i_{j,k'})=\textsf{A}$}
\State $\textup{viol}(y)^\textsf{p}_{\mathsf{B},\mathsf{A}}.\textup{add}(\varsigma^i_{j,k})$
\EndIf
\If{$\not\exists {h,k'}.\;\lambda(\varsigma^i_{h,k'})=\textsf{B} \textbf{\,and\,} j+{\color{blue}(\texttt{poly}\textbf{\,\textrm{?}\,}\pi(\varsigma^i_{j,k})\textrm{\textbf{\,:\,}}1)}{\color{blue}\leq} h$}\label{ignoreSpan2}
\State $\textup{viol}(y)^\textsf{r}_{\mathsf{A},\mathsf{B}}.\textup{add}(\varsigma^i_{j,k})$
\Else \, ${\textup{act}(y)}^\textsf{r}_{\mathsf{A},\mathsf{B}}.\textup{add}(\braket{\varsigma^i_{j,k},\varsigma^i_{h,k'}})$ \textbf{for all}  $h,k'$ \textbf{s.t.} $\lambda(\varsigma^i_{h,k'})=\textsf{B} \textbf{\,and\,} j+{\color{blue}(\texttt{poly}\textbf{\,\textrm{?}\,}\pi(\varsigma^i_{j,k})\textrm{\textbf{\,:\,}}1)}{\color{blue}\leq} h$
 
\EndIf

\EndIf

\EndFor
\EndIf

\EndProcedure

\end{algorithmic}
\end{spacing}
\end{algorithm}

Algorithm \ref{algo:collectEvidence} extends the mechanism in Bolt2 \cite{bolt2} and used in the previous paper for efficiently mining the aforementioned clauses by shifting from the collection of the trace IDs activating (\textsf{sat}), violating (\textsf{viol}), or not activating (\textsf{vac}) the clauses, to retaining which constituents precisely generate these conditions. This is performed to reconstruct the payloads associated with the activated and targeted conditions, and to reconstruct which constituent led to the violation of the clause. We denote all the possible target conditions for each activation leading to clause satisfaction as pairs of constituents associated with the activation. Bolt2  (e.g., Line \ref{ignoreSpan1} and Line \ref{ignoreSpan2}) did not consider different overlapping constituents referring to the same \gls{mts} variable to avoid redundant and obvious correlations. To recognize those, we now use activity label taxonomies: for constituents expressing distinct \gls{dt} associated with the same \gls{mts} variable, we are interested in establishing temporal correlations between activated and targeted constituents if and only if the targeted (or activated) constituent terminates before the occurrence of the forthcoming activated (or targeted) one while always referring to the same variable. Bolt2 only considered pointwise non-polyadic events with no associated durative information, thus proving to be inadequate to support these new features.

\section{Empirical Results}

We exploit some time series datasets made available through the \texttt{sktime} library \cite{sktime} while considering the Dyskinetic event dataset from our previous contribution \cite{ideas2024a}. Italy Power Demand \cite{italypowerdemand} provides a dataset of univariate and equal-length time series, where the classification label is used to differentiate energy consumption patterns in the October-March period vs the April-September ones. The Basic Motions dataset\footnote{\url{http://timeseriesclassification.com/description.php?Dataset=BasicMotions}} considers multivariate time series with 6 dimensions and distinguishing different motion types, walking, resting, running, and badminton. This dataset collects the $x$, $y$, and $x$ axis information from wrist-watch accelerometer and a gyroscope; this dataset has been used to compare this motion dataset with the one discussed in our previous work while remarking the inherent more difficult nature of solving this other clinical problem.
 The Dyskinetic dataset \cite{ideas2024a} attempts at categorize Dyskinetic/Off events in terms of different drug assumption patterns as well as motor sensors information. Differently from the previous set of experiments \cite{ideas2024a}, we now consider for both our algorithms and the competing \gls{mtsc} approaches the full dataset also considering the active principles assumption intakes estimated using a rough approximation from current literature \cite{Arav2023}. We also use the OSULeaf \cite{osuleaf} dataset to remark how a simple image classification problem can be also represented as a time series classification problem: the univariate series were obtained by color image segmentation and boundary extraction (in the anti-clockwise direction) from digitized leaf images of six classes: Acer Circinatum, Acer Glabrum, Acer Macrophyllum, Acer Negundo, Quercus Garryana, and Quercus Kelloggii.

We consider such datasets for the following reasons: despite \gls{emeritate} was explicitly design to capture trend correlations across  different dimensions, we also use univariate datasets such as Italy Power Demand and OsuLeaf to show that this approach can be also applied to simpler dataset. Furthermore, we consider another multivariate dataset also considering different motor sensors to showcase the peculiar difference between the problem posed by the identification of the Dyskinetic Events from simply detecting different movement patterns. We provide some general statistics for these  in Table \ref{tab:stats}.

\begin{table}[!h]
\caption{Dataset statistics}\label{tab:stats}
\begin{tabular}{lrrrrrr}
\toprule
Dataset & \#MTS  & $\textit{avg}_T$ Segments & \#Dimensions   & $\textit{avg}_T|T|$ &  $\max_T\Sigma_T$ & \#Classes\\
\midrule
\textit{Italy Power Demand} & 1096   & 1  & 1  &  24 & 13& 2\\
\textit{Basic Motions} 		& 80     & 1  & 6  &  100& 72& 4\\
\textit{OsuLeaf} 			& 442    & 1  & 1  &  427& 12& 6\\
\textit{Dyskinetic Events} 	& 2      & 23 & 30 & 44  & 436& 2\\
\bottomrule
\end{tabular}
\end{table}

Our benchmarks were conducted on a Dell Precision mobile workstation 5760 running Ubuntu 22.04. The specifications of the machine include an Intel\textsuperscript{\textregistered} Xeon(R) W-11955M CPU @ 2.60GHz $\times$ 16, 64GB DDR4 3200MHz RAM, with 500 GB of free disk space.

\subsection{Run-time efficiency of \gls{emeritate} vs \gls{emeritatedf}}
\figurename~\ref{fig:finegrain} refers to the running times of both our proposed \gls{mtsc} algorithms, from both our previous contribution and from the current paper. The first three phases refer to the A Priori explainability segment, while the last two refer to the Ad Hoc one. In both scenarios, we consider $\theta=0$ for maximising the DECLARE recall, and consider in both circumstances a sensitivity parameter of $\varepsilon=10^{-4}$. We considered the time required to fully process each dataset.

\begin{figure}[!h]
\includegraphics[width=\linewidth]{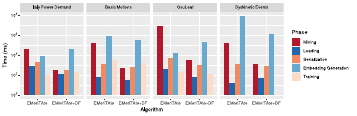}
\caption{Fine-Grained Mining+Learning times for the training phase of both \gls{emeritate} and \gls{emeritatedf}.}\label{fig:finegrain}
\end{figure}

Although the loading times of the datasets are comparable, we observe that the loading phase in this novel solution, which is different from the former, also includes a time series indexing phase. This then leads to a significant improvement in time minimisation, also ascribable to discarding some redundant \gls{dt} activity labels. This is also witnessed in the serialization phase, which has slightly decreased despite the addition of Catch22 payloads. Running times for the mining phase in \gls{emeritatedf} remark that this algorithm is heavily dominated by the number of traces and segments being available as, in these situations, the former version of the algorithm was heavily hampered by conducting the mining and the embedding phases in two different times, thus requiring to load the data and index it several times. The newly proposed approach might have a more significant overhead in some situations. In fact, generating a dataful representation of the clauses requires accessing the constituents' payloads several times, thus increasing the number of times we access KnoBAB for retrieving log information. Notwithstanding the former, this leads to a decrease in running time (\figurename~\ref{allEfficiency}).

\begin{figure}[!h]
\includegraphics[width=\linewidth]{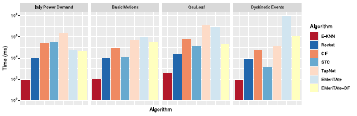}
\caption{Comparing the cumulative training times.}\label{allEfficiency}
\end{figure}

\subsection{Comparing competing \gls{mtsc} approaches}

We used the Python \texttt{sktime} library to ensure a uniform implementation for all competing approaches discussed in Section \ref{competingApproaches}, for which we used the default hyperparameters setup from \texttt{sktime}. As all of the competing classifiers do not support classification of time series of different length, and given that these classifiers do not support time series varying their classification outcome in time, for the sole Dyskinetic event dataset, we consider the maximal projections belonging to the same class and, between each of these, we consider multiple sliding windows of the same size. This overall builds up the number of time series to be classified by competing approaches.

\subsubsection{Run-time efficiency.} \figurename~\ref{allEfficiency}
 compares the running time for all the \gls{mtsc} approaches being trained over all the aforementioned datasets without splitting the data into training and testing datasets, thus using the same setting for the previous experiments. Competing approaches are not significantly impacted over the number of \gls{mts} dimensions as our approaches: please observe the tunning times for the Dyskinetic events where, despite the number of traces increased so to obtain multiple \gls{mts} of the same length, the competing approaches still provide a lesser running time if compared to our approaches. This is motivated by our solutions testing multiple possible correlations across dimension and activity labels, while other solutions such as Rocket and TapNet, attempt to capture cross-dimensional correlations through convolutions, thus flattening up the values via dimensionality reduction. Similar considerations can also be carried out by the other classifiers, which have not been specifically designed  to consider correlations across dimensions: Euclidean distance for \gls{mts} only considers dimensionality-wise similarity, CIF considers variations within one fixed time and length sliding window while considering variations only within one single dimension, and STC considers shapelets only occurring within one single dimension. On the other hand, TapNet is heavily dominated by the number of traces being considered at training time and, across all the competitors, is the one always requiring a considerable amount on training time. \gls{emeritate} solution overtakes \gls{emeritatedf} over the Dyskinetic dataset, while the other dataset consistently shows similar running times.

\begin{figure}
\centering
\begin{subtable}{1\textwidth}
\begin{tabular}{l r@{\hspace*{2mm}}r@{\hspace*{2mm}}r@{\hspace*{2mm}}r}
\toprule
 \textit{Algorithm}                   & Accuracy (\%)    & Precision (\%)   & Recall (\%)      & F1 (\%)          \\
\midrule
 E-KNN                & $96.47 \pm \,\,1.22$ & $ {\boldmath 97.36 \pm\,\, 1.19}$ & $\color{red}95.58 \pm\,\, 1.82$ & $96.45 \pm\,\, 1.24$ \\
 Rocket  \cite{DempsterPW20}                     & ${97.02 \pm \,\,1.06}$ & $96.70 \pm\,\, 1.72$  & ${97.39 \pm\,\, 1.21}$ & ${97.04 \pm \,\,1.05}$ \\
 CIF \cite{9378424}     & $96.69 \pm \,\,1.82$ & $96.73 \pm\,\, 1.75$ & $96.67 \pm\,\, 3.03$ & $96.69 \pm\,\, 1.87$ \\
 STC \cite{Hills2014} & $96.81 \pm \,\,1.37$ & $96.81 \pm\,\, 1.70$  & $96.85 \pm\,\, 3.03$ & $96.81 \pm\,\, 1.42$ \\
 TapNet \cite{ZhangG0L20}                      & $\color{red}96.20 \pm \quad\,\,\,\,\varepsilon$ & $\color{red}94.76 \pm\,\, 1.62$ & $97.88 \pm\,\, 0.91$ & $96.28 \pm \quad\,\,\,\,\varepsilon$ \\

\hdashline[0.5pt/2pt]
\textbf{\color{green}EMeriTAte} \cite{ideas2024a} & $\color{blue}99.59 \pm\quad\,\,\,\,\varepsilon$ & $\color{blue}99.73 \pm\quad\,\,\,\,\varepsilon$ & $\color{blue}99.45 \pm\,\, 0.01$ & $\color{blue}99.59 \pm \quad\,\,\,\,\varepsilon$\\
\textbf{EMeriTAte+DF} &  $96.23\pm\,\,1.52$ & $96.24\pm\,\,1.51$ & $96.23\pm\,\,1.52$ & $\color{red}96.23\pm\,\,1.52$ \\
\bottomrule
\end{tabular}
\caption{Italy Power Demand}
\end{subtable}
\begin{subtable}{1\textwidth}
\begin{tabular}{lrr@{\hspace*{2mm}}r@{\hspace*{2mm}}r}
\toprule
 \textit{Algorithm}                   & Accuracy (\%)    & M. Precision (\%)   & M. Recall (\%)   & M. F1 (\%)    \\
\midrule
 E-KNN                & $\color{red}62.50 \pm 8.33$  & $\color{red}53.09 \pm 3.79$       & $\color{red}62.50 \pm 8.33$     & $\color{red}53.91 \pm 7.15$ \\
 Rocket \cite{DempsterPW20}                     & $99.58 \pm 2.08$ & $99.64 \pm 1.79$       & $99.58 \pm 2.08$    & $99.58 \pm 2.10$  \\
 CIF  \cite{9378424}   & $\color{blue}{100.0 \pm 0.00}$ & $\color{blue}{100.0 \pm 0.00}$       & $\color{blue}{100.0 \pm 0.00}$    & $\color{blue}{100.0 \pm 0.00}$ \\

 STC \cite{Hills2014} & $99.58 \pm 2.08$ & $99.64 \pm 1.79$       & $99.58 \pm 2.08$    & $99.58 \pm 2.10$  \\
 TapNet \cite{ZhangG0L20}                          & $\color{blue}{100.0 \pm 0.00}$ & $\color{blue}{100.0 \pm 0.00}$       & $\color{blue}{100.0 \pm 0.00}$    & $\color{blue}{100.0 \pm 0.00}$ \\

\hdashline[0.5pt/2pt]
\textbf{EMeriTAte} \cite{ideas2024a} & $91.66\pm 8.33$&$92.69\pm 7.81$ & $91.66\pm 8.33$ & $91.51\pm 8.57$ \\
\textbf{\color{green}EMeriTAte+DF} & $98.75\pm 2.08$ & $98.92\pm 1.78$ & $98.75\pm2.08$ & $98.74\pm2.10$ \\
\bottomrule
\end{tabular}
\caption{Basic Motions}
\end{subtable}
\begin{subtable}{1\textwidth}
\begin{tabular}{lrrr@{\hspace*{2mm}}r}
\hline
 \textit{Algorithm}    & Accuracy (\%)    & M. Precision (\%)   & M. Recall (\%)   & M. F1 (\%)    \\
\hline
 E-KNN & $\color{red}63.46 \pm 6.02$ & $\color{red}65.43 \pm 5.79$       & $\color{red}62.59 \pm 6.53$    & $\color{red}62.98 \pm 6.54$ \\
 Rocket  \cite{DempsterPW20}     & $96.32 \pm 2.26$ & $96.77 \pm 1.98$       & $95.98 \pm 2.41$    & $96.25 \pm 2.24$ \\
CIF \cite{9378424} & $84.44 \pm 3.01$ & $86.55 \pm 3.25$       & $82.38 \pm 4.94$    & $83.29 \pm 4.55$ \\
STC \cite{Hills2014} & $92.86 \pm 3.01$ & $93.35 \pm 3.23$       & $91.72 \pm 3.78$    & $92.25 \pm 3.56$ \\
TapNet  \cite{ZhangG0L20}     & $82.71 \pm 1.50$  & $88.29 \pm 1.06$       & $78.56 \pm 5.49$    & $78.59 \pm 6.65$ \\ 
\hdashline[0.5pt/2pt]
\textbf{EMeriTAte}\cite{ideas2024a} & $79.62\pm 8.27$ & $83.72\pm7.78$ & $80.13\pm8.50$ & $80.39\pm8.56$ \\
\textbf{\color{green}EMeriTAte+DF} & $\color{blue}99.70\pm0.01$ & $\color{blue}99.63\pm0.01$ & $\color{blue}99.70\pm\,\,\,\,\,\,\,\, \varepsilon$ & $\color{blue}99.65\pm\,\,\,\,\,\,\,\, \varepsilon$ \\
\hline
\end{tabular}
\caption{OsuLeaf}
\end{subtable}
\begin{subtable}{1\textwidth}
\begin{tabular}{ll@{\hspace*{2mm}}r@{\hspace*{2mm}}r@{\hspace*{2mm}}r}
\toprule
 \textit{Algorithm}    & Accuracy (\%)    &  Precision (\%)   &  Recall (\%)   &  F1 (\%)    \\
\midrule
 E-KNN                & $\color{red}12.14 \pm 10.71$ & $10.06 \pm 18.18$ & $12.86 \pm 28.57$ & $11.21 \pm 22.22$ \\
 Rocket  \cite{DempsterPW20}                    & $\color{red}12.14 \pm 10.71$ & $7.80 \pm 18.18$   & $10.00 \pm 28.57$  & $8.65 \pm 22.22$  \\
 CIF \cite{9378424}    & $\color{red}12.14 \pm 10.71$ & $\color{red}4.58 \pm 16.67$  & $\color{red}5.71 \pm 21.43$  & $\color{red}5.08 \pm 18.75$  \\
 STC \cite{Hills2014} & $25.71 \pm 17.86$ & $14.44 \pm 22.22$ & $12.86 \pm 28.57$ & $13.02 \pm 25.00$  \\ 
 TapNet  \cite{ZhangG0L20}     & $15.00 \pm 10.71$  & $15.19 \pm 18.18$ & $20.0 \pm 28.57$  & $17.09 \pm 22.22$ \\ 

\hdashline[0.5pt/2pt]
\textbf{EMeriTAte} \cite{ideas2024a} & ${98.57 \pm\,\,\, 7.14}$ & $\color{blue}{100.0 \pm\,\,\, 0.00}$       & ${97.14 \pm 14.28}$    & ${98.33 \pm \,\,\, 8.33}$ \\
\textbf{\color{green}EMeriTAte+DF} & $\color{blue}{100.0 \pm\,\,\, 0.00}$ & $\color{blue}{100.0 \pm \,\,\,0.00}$       & $\color{blue}{100.0 \pm\,\,\, 0.00}$    & $\color{blue}{100.0 \pm\,\,\, 0.00}$ \\
\bottomrule
\end{tabular}
\caption{Dyskinetic Events}
\end{subtable}
\caption{Training Results over the datasets of interest. Macro metrics are used for datasets containing more than 2 classes. Numbers in blue (red) remark the best (worst) results.}\label{classout}
\end{figure}
\subsubsection{Classification outcomes.}
\figurename~\ref{classout} provides the accuracy, precision, recall, and F1 score results across all the aforementioned datasets and competing approaches when testing the solutions with a 70\%-30\% split between training and testing dataset using stratified k-fold sampling. For both \gls{emeritate} and \gls{emeritatedf}, we consider training the Decision Trees over the trace embeddings with a maximum depth of $5$, so to minimise the chance of generating models overfitting the data. We run 10 experiments, for which we report the average score and the distance between the greatest and the lowest value obtained ($\pm$). When considering datasets with more than two classes, we use macro measures. Concerning the Dyskinetic dataset, we kept different training/testing splits from the ones in our previous paper. 

We observe that, for three datasets out of four, \gls{emeritatedf} outperforms our previous implementation. The major score distance between these two solutions can be found on the OsuLeaf dataset: this can be explained as the main difference between the two solutions resides in the usage of the data refinement, while the discard of \gls{dt} patterns seems not to have interfered with the classification outcome. For the Italy Power Demand dataset,  we see that our first solution outperforms the current implementation: given the above, this can be ascribed to the presence of several fluctuations events which were not discarded in our previous solution. 

By comparing these solutions to the competing approaches, no competitor can constantly outperform the others across all the datasets and, when \gls{emeritatedf} did not achieve maximum scores, it still achieved $\approx 99\%$ of accuracy, macro precision, macro recall, and F1 score. E-KNN scores low values for datasets such as the former and OsuLeaf, thus remarking the impracticality of merely considering time series similarity as a valid pathway for classification outcomes for real-world datasets. 

While comparing the results over the Dyskinetic dataset, it is clear that the best approaches were the ones considering silhouette (STC) and attention ones (TapNet) while still scoring metrics way below $50\%$. This can be motivated as follows: either the process of splitting the time series lost some important correlation information or truncated some patterns, or these solutions still cannot account for extracting the best behavioural predictors for the dataset of interest. After analysing the models extracted from \gls{emeritatedf}, we can clearly see that the classification always refers to correlations across different dimensions. Given the explanation being extracted, this then remarks that the dataset of interest is heavily characterized by correlations across dimensions that could not be adequately captured by the pre-existing solutions. 

\section{Conclusion and Future Works}
This paper provides the first algorithm towards a generalization of existing algorithms in the literature for classifications of temporal behaviours: in fact, we use a purely event-driven technique to classify multivariate numerical temporal sequences. By doing so, we have shown how it is possible to appropriately discretize a numerical representation, maintaining as much of the information in the data as possible. To appropriately characterize the numerical data, we have also deemed it necessary to further generalize the representation of temporal data to events, prescribing that multiple distinct events of different durations can occur at a given instant of time. From the generalization of the deviance mining algorithms, we obtain a new classifier that can be used to classify numerical time series with good results in precision, recall, and accuracy. Subsequent studies will also establish this mechanism's validity in characterizing event-based data.

Future works will investigate whether considering additional events not currently mined in the indexed \gls{dt} phase will make \gls{emeritatedf} reach the same precision, recall, and accuracy results from our previous contribution, \gls{emeritate}, without tampering the running time of the subsequent \textit{ad hoc} phase. Despite preliminary results remark that the non-negligible running time of this latter phase might be due to the considerable presence of events to ascertain, considering larger amounts of data will necessarily require further improving over the presently proposed algorithms. As both our solutions are implemented using KnoBAB as a temporal-driven database, future works will also require to extend it so to move from a in-memory data representation to a one using secondary memory.
We will also consider the possibility of applying the general-to-specific mining approach that we previously investigated over Bolt2 for mining \gls{dt} patterns, so to achieve mining efficiency without resorting to discarding noisy or redundant events.

Experiments over these datasets lead dataful specifications describing the classes showing only refinements over the activation conditions, but not on the target ones. We will also test the possibility of other datasets also requiring the definition of  correlation conditions between activated and targeted events, thus requiring the definition of comparisons between activated and targeted constituents' payloads. Despite the possibility of doing so though weighted oblique trees \cite{Yang_Shen_Gao_2019}, this decision will come at the cost of significantly adding computational overhead over the refinement phase. As a byproduct of this, further work needs also to be done to further improve over the fitting of oblique decision trees.

\bibliographystyle{splncs03}
\bibliography{example}

\begin{thebibliography}{10}
\providecommand{\url}[1]{\texttt{#1}}
\providecommand{\urlprefix}{URL }

\bibitem{vanderAalst2022}
van~der Aalst, W.M.P.: Discovering Directly-Follows Complete Petri Nets
  from Event Data, pp. 539--558. Springer Nature Switzerland, Cham (2022)

\bibitem{XES}
Acampora, G., Vitiello, A., Di~Stefano, B., van~der Aalst, W., Gunther, C.,
  Verbeek, E.: {IEEE 1849: The XES Standard: The Second IEEE Standard Sponsored
  by IEEE Computational Intelligence Society}. IEEE Comp. Int. Mag.  12(2)
  (2017)

\bibitem{Allen}
Allen, J.F.: Maintaining knowledge about temporal intervals. Commun. ACM
  26(11),  832–843 (Nov 1983)

\bibitem{wrembel}
Andrzejewski, W., Bębel, B., Boiński, P., Wrembel, R.: On tuning parameters
  guiding similarity computations in a data deduplication pipeline for
  customers records: Experience from a r\&d project. Information Systems  121,
  102323 (2024),
  \url{https://www.sciencedirect.com/science/article/pii/S030643792300159X}

\bibitem{5963680}
Anselma, L., Bottrighi, A., Montani, S., Terenziani, P.: Extending bcdm to cope
  with proposals and evaluations of updates. IEEE Transactions on Knowledge and
  Data Engineering  25(3),  556--570 (2013)

\bibitem{Arav2023}
Arav, Y., Zohar, A.: Model-based optimization of controlled release formulation
  of levodopa for parkinson's disease. Scientific Reports  13(1),  15869 (Sep
  2023)

\bibitem{AvolioFVZ23}
Avolio, M., Fuduli, A., Vocaturo, E., Zumpano, E.: On detection of diabetic
  retinopathy via multiple instance learning. In: Proceedings of the
  International Database Engineered Applications Symposium Conference, {IDEAS}
  2023, Heraklion, Crete, Greece, May 5-7, 2023. pp. 170--176. {ACM} (2023)

\bibitem{202403.0286}
Bergami, G.: {DECLARE}$d$: {A} {Polytime} {LTL}\textsubscript{f} {Fragment}.
  Logics  2(2),  79--111 (2024)

\bibitem{bolt2}
Bergami, G., Appleby, S., Morgan, G.: Specification mining over temporal data.
  Computers  12(9) (2023)

\bibitem{10.1145/3410566.3410583}
Bergami, G., Bertini, F., Montesi, D.: Hierarchical embedding for dag
  reachability queries. In: Proceedings of the 24th Symposium on International
  Database Engineering \& Applications. IDEAS '20, Association for Computing
  Machinery, New York, NY, USA (2020)

\bibitem{chapter}
Bergami, G., Fox, O.R.: Extracting specifications through verified and
  explainable ai: Interpretability, interoperability, and trade-offs. Logics  3
  (2025 (To Appear))

\bibitem{Undone}
Bergami, G., Francescomarino, C.D., Ghidini, C., Maggi, F.M., Puura, J.:
  Exploring business process deviance with sequential and declarative patterns.
  CoRR  abs/2111.12454 (2021)

\bibitem{9576856}
Bergami, G., Maggi, F.M., Montali, M., Peñaloza, R.: Probabilistic trace
  alignment. In: 2021 3rd International Conference on Process Mining (ICPM).
  pp. 9--16 (2021)

\bibitem{ideas2024a}
Bergami, G., Packer, E., Scott, K., Del~Din, S.: Predicting dyskinetic events
  through verified multivariate time series classification. In: Database
  Engineered Applications. IDEAS '24 (in press), Springer (2025)

\bibitem{XuLZ17a}
{De Giacomo}, G., Maggi, F.M., Marrella, A., Patrizi, F.: {On the Disruptive
  Effectiveness of Automated Planning for LTL\emph{f}-Based Trace Alignment}.
  In: AAAI'17. {AAAI} press (2017)

\bibitem{DempsterPW20}
Dempster, A., Petitjean, F., Webb, G.I.: {ROCKET:} exceptionally fast and
  accurate time series classification using random convolutional kernels. Data
  Min. Knowl. Discov.  34(5),  1454--1495 (2020)

\bibitem{donze13}
Donz{\'{e}}, A.: On signal temporal logic. In: Runtime Verification - 4th
  International Conference, {RV} 2013, Rennes, France, September 24-27, 2013.
  Proceedings. Lecture Notes in Computer Science, vol. 8174, pp. 382--383.
  Springer (2013)

\bibitem{osuleaf}
Gandhi, A.: Content-Based Image Retrieval: Plant Species Identification.
  Master's thesis, Oregon State University (2002)

\bibitem{Hills2014}
Hills, J., Lines, J., Baranauskas, E., Mapp, J., Bagnall, A.: Classification of
  time series by shapelet transformation. Data Mining and Knowledge Discovery
  28(4),  851--881 (Jul 2014)

\bibitem{HUO2022117176}
Huo, X., Hao, K., Chen, L., song Tang, X., Wang, T., Cai, X.: A dynamic soft
  sensor of industrial fuzzy time series with propositional linear temporal
  logic. Expert Systems with Applications  201,  117176 (2022)

\bibitem{Inoue2024}
Inoue, T., Kubota, K., Ikami, T., Egami, Y., Nagai, H., Kashikawa, T., Kimura,
  K., Matsuda, Y.: Clustering method for time-series images using
  quantum-inspired digital annealer technology. Communications Engineering
  3(1), ~10 (Jan 2024), \url{https://doi.org/10.1038/s44172-023-00158-0}

\bibitem{italypowerdemand}
Keogh, E., Wei, L., Xi, X., Lonardi, S., Shieh, J., Sirowy, S.: Intelligent
  icons: Integrating lite-weight data mining and visualization into gui
  operating systems. In: Sixth International Conference on Data Mining
  (ICDM'06). pp. 912--916 (2006)

\bibitem{LENO2020101482}
Leno, V., Dumas, M., Maggi, F.M., {La Rosa}, M., Polyvyanyy, A.: Automated
  discovery of declarative process models with correlated data conditions.
  Information Systems  89,  101482 (2020)

\bibitem{sktime}
L{\"{o}}ning, M., Bagnall, A.J., Ganesh, S., Kazakov, V., Lines, J.,
  Kir{\'{a}}ly, F.J.: sktime: {A} unified interface for machine learning with
  time series. CoRR  abs/1909.07872 (2019),
  \url{http://arxiv.org/abs/1909.07872}

\bibitem{catch24}
Lubba, C.H., Sethi, S.S., Knaute, P., Schultz, S.R., Fulcher, B.D., Jones,
  N.S.: catch22: Canonical time-series characteristics - selected through
  highly comparative time-series analysis. Data Min. Knowl. Discov.  33(6),
  1821--1852 (2019)

\bibitem{9378424}
Middlehurst, M., Large, J., Bagnall, A.: The canonical interval forest (cif)
  classifier for time series classification. In: 2020 IEEE International
  Conference on Big Data (Big Data). pp. 188--195 (Dec 2020)

\bibitem{PesicSA07}
Pešić, M., Schonenberg, H., van~der Aalst, W.M.P.: {DECLARE:} full support
  for loosely-structured processes. In: {EDOC}. pp. 287--300. {IEEE} Computer
  Society (2007)

\bibitem{RostGTFSCAJR22}
Rost, C., G{\'{o}}mez, K., T{\"{a}}schner, M., Fritzsche, P., Schons, L.,
  Christ, L., Adameit, T., Junghanns, M., Rahm, E.: Distributed temporal graph
  analytics with {GRADOOP}. {VLDB} J.  31(2),  375--401 (2022)

\bibitem{Ruiz2021}
Ruiz, A.P., Flynn, M., Large, J., Middlehurst, M., Bagnall, A.: The great
  multivariate time series classification bake off: a review and experimental
  evaluation of recent algorithmic advances. Data Mining and Knowledge
  Discovery  35(2),  401--449 (Mar 2021)

\bibitem{10.1145/3503914}
Seshia, S.A., Sadigh, D., Sastry, S.S.: Toward verified artificial
  intelligence. Commun. ACM  65(7),  46–55 (Jun 2022),
  \url{https://doi.org/10.1145/3503914}

\bibitem{Yang_Shen_Gao_2019}
Yang, B.B., Shen, S.Q., Gao, W.: Weighted oblique decision trees. Proceedings
  of the AAAI Conference on Artificial Intelligence  33(01),  5621--5627 (Jul
  2019), \url{https://ojs.aaai.org/index.php/AAAI/article/view/4505}

\bibitem{ZhangG0L20}
Zhang, X., Gao, Y., Lin, J., Lu, C.: {TapNet}: Multivariate time series
  classification with attentional prototypical network. In: AAAI. pp.
  6845--6852. {AAAI} Press (2020)

\end{thebibliography}

\end{document}